\documentclass{amsart}

\usepackage{graphicx}
\usepackage{makecell}
\newtheorem{theorem}{Theorem}[section]

\usepackage{hyperref}
\theoremstyle{definition}
\newtheorem{definition}[theorem]{Definition}

\theoremstyle{remark}

\numberwithin{equation}{section}

\usepackage{amssymb}
\begin{document}

\title{Formal structure of periodic system of elements}

\author{Wilmer Leal}
%%%%%%%%% Insert author address here
\address{Bioinformatics Group, Department of Computer Science, Leipzig University, Germany}
\curraddr{Max Planck Institute for Mathematics in the Sciences, Leipzig, Germany}
\email{wleal@mis.mpg.de}
\thanks{WL was supported by a PhD scholarship from the \emph{German Academic Exchange
    Service} (DAAD): Forschungsstipendien-Promotionen in Deutschland, 2017/2018 (Bewerbung 57299294)}

\author{Guillermo Restrepo}
\address{Max Planck Institute for Mathematics in the Sciences, Leipzig, Germany}
\curraddr{Interdisciplinary Center for Bioinformatics, Leipzig University, Germany}
\email{restrepo@mis.mpg.de}
\thanks{GR thanks the  support of the Alexander von Humboldt Foundation}

%    General info
%\subjclass[2000]{Primary 05C65, 14E20; Secondary 46E25, 20C20}

%\date{February 1, 2001}



\begin{abstract}
For more than 150 years the structure of the periodic system of the chemical elements has intensively motivated research in different areas of chemistry and physics.  However, there is still no unified picture of what a periodic system is.  Herein, based on the relations of order and similarity, we report a formal mathematical structure for the periodic system, which corresponds to an ordered hypergraph.  It is shown that the current periodic system of chemical elements is an instance of the general structure.  The definition is used to devise a tailored periodic system of polarizability of single covalent bonds, where order relationships are quantified within subsets of similar bonds and among these classes.  The generalised periodic system allows envisioning periodic systems in other disciplines of science and humanities.
\end{abstract}

\maketitle

\section{Introduction}
%%%% Insert A head here
Since the formulation of the periodic system in the 1860s, the quest for understanding its structure has intensively motivated research in different areas of chemistry and physics.  However, almost 150 years after its announcement, the different approaches from quantum chemistry \cite{Schwarz2010a, Schwarz2013, Huang2016, Pyykko2011, Pyykko2017, Pyykko2016, Pyykko2012, Schwarz2010b, Wahiduzzaman2013, Geerlings2011, Schwarz2009}, group theory \cite{Thyssen2013, Dudek2002}, clustering \cite{Schwarz2009, Leal, RestrepoOUP2} and information theory \cite{Bonchev, Geerlings2011}, to name but a few \cite{mathPTbook,Imyanitov2011}, have not led to an unified picture \cite{Katriel2012}.  Instead, they give insights on the possible chemical and physical causes of the patterns depicted by the system but have failed in providing a formal structure for it \cite{Katriel2012}.

As noted by Mendeleev:\footnote{According to Scerri \cite{Scerri2007}, Mendeleev was one of the six formulators of the periodic system, being the others B\'eguyer de Chancourtois, Newlands, Meyer, Odling and Hinrichs.} ``the reason for the absence of any explanation concerning the nature of the periodic law [Here, in general, periodic system] resides entirely in the fact that not a single rigorous, abstract expression of the law has been discovered (p. 221 of reference \cite{Jensen})." In this paper we report a formal structure for periodic system, based on a contemporary mathematical interpretation of 1869 Mendeleev publication and recent studies of the system.

\subsection{Periodic system, table and periodic law}

These are different terms that are usually treated as synonyms, but even if related, they make reference to different subjects \cite{RestrepoOUP2}.  For the sake of clarity, here we discuss their differences.

A \emph{periodic system} of chemical elements is the structure resulting from considering order and similarity of chemical elements.  A \emph{periodic table} is a mapping of the periodic system to another space, normally a bi-dimensional space.  By \emph{periodic law} is understood the observed oscillation of some properties of chemical elements as a function of the atomic number $Z$.

There is not only one periodic system for the chemical elements, for they depend on the considered elements and on the setting up of similarity and order.  Likewise, the intended generality of \emph{the} periodic law to \emph{all} properties of chemical elements does not hold, for there are properties that do not oscillate with $Z$.

In the current paper we explore the structure of a periodic system.

\subsection{The role of similarity and order}

In his 1869 publication, Mendeleev wrote: ``if one arranges the elements in vertical columns according to increasing atomic weight, such that the horizontal rows contain analogous elements, also arranged according to increasing atomic weight, one obtains the following table" (p. 16 of \cite{Jensen}). After considering that current tables interchange Mendeleev's columns and rows and that the ``arranging'' criteria has been replaced by the atomic number, two important relations are the salient structure keepers of the table, and in general of the periodic system: \emph{order} and \emph{similarity}.

Before going any further, let us analyse these two relations through examples.  Let us take H, He and Li and their atomic numbers, which we order with the usual order on natural numbers, denoted by $\preceq$.  An order relation holds that every element is related to itself, e.g. H $\preceq$ H.  It holds that if He $\preceq x$ and if $x \preceq$ He, then $x$ is He. In addition, if H $\preceq$ He, and He $\preceq$ Li, then H $\preceq$ Li.  In short, \emph{an order relation is reflexive, antisymmetric and transitive} (Appendix-Definition \ref{order}).  If $E$ is the set of elements, its order by $\preceq$ is denoted as $(E,\preceq)$.  In contrast to order, \emph{similarity}, represented as $\sim$, \emph{is only reflexive and symmetric}, that is, self similarity is allowed (Na $\sim$ Na) and if Na is similar to K (Na $\sim$ K), then K $\sim$ Na (Appendix-Definition \ref{similarity}).  As similarity is used for classifying, it is worth mentioning that a customary outcome of a classification is a partition (Appendix-Definition \ref{partition}), i.e. a collection of subsets not sharing elements.  The suitability of partitions for periodic systems is discussed later.

Despite the relevance of similarity and order for the periodic system \cite{Ashcroft2017},\footnote{Nicely accounted by the Oxford Dictionary when referring to the periodic table as: ``A table of the chemical elements arranged in order of atomic number, usually in rows, so that elements with similar atomic structure (and hence similar chemical properties) appear in vertical columns''\cite{Oxford}.} they are considered as separate aspects of it, with some emphasis on classification \cite{Scerri2009,Bengoetxea2014,Scerri2012a,Scerri2012b}\footnote{An example from outside the scientific literature is the definition of periodic table by the Cambridge Dictionary: ``An arrangement of the symbols of chemical elements in rows and columns, showing similarities in chemical behaviour, especially between elements in the same columns'' \cite{Cambridge}.} caused by the, taken for granted, ordering of the elements based upon atomic number.  Whereas the possibilities for classifying are multiple given the huge number of properties chemical elements have, an exceptional example stressing ordering over similarity for the case of the table is the definition from Wikipedia: ``The periodic table is a tabular arrangement of the chemical elements, ordered by their atomic number, electron configurations, and recurring chemical properties'' \cite{Wikipedia}.  In some other cases, as in \cite{Ashcroft2017,Scerri7tale}, it is said that similarity begets ordering, while, for example, the Encyclopaedia Britannica states the opposite: ``the organized array of all the chemical elements in order of increasing atomic number i.e., the total number of protons in the atomic nucleus. When the chemical elements are thus arranged, there is a recurring pattern called the `periodic law' in their properties, in which elements in the same column (group) have similar properties''\cite{Encyclopaediabritannica}. 

Hence, there is confusion between order and similarity (classification), which are different binary relations, and the confusion has led to wrong statements that order leads to classifications and the other way round.  The distinction between these two relations is central for the structure of the periodic system.

\section{The structure of the periodic system}

\subsection{Mendeleevian periodic system}

Back to Mendeleev's statement, the ``ingredients'' of a periodic system are: chemical elements ($E$), order by atomic number ($\preceq_Z$) and a classification ($C_P$) of the elements based on some properties $P$.\footnote{Although here we imply an unsupervised classification, it may be supervised too.}

\begin{figure}[h]
	\centering
	\includegraphics[width=.9\textwidth,height=!, keepaspectratio]{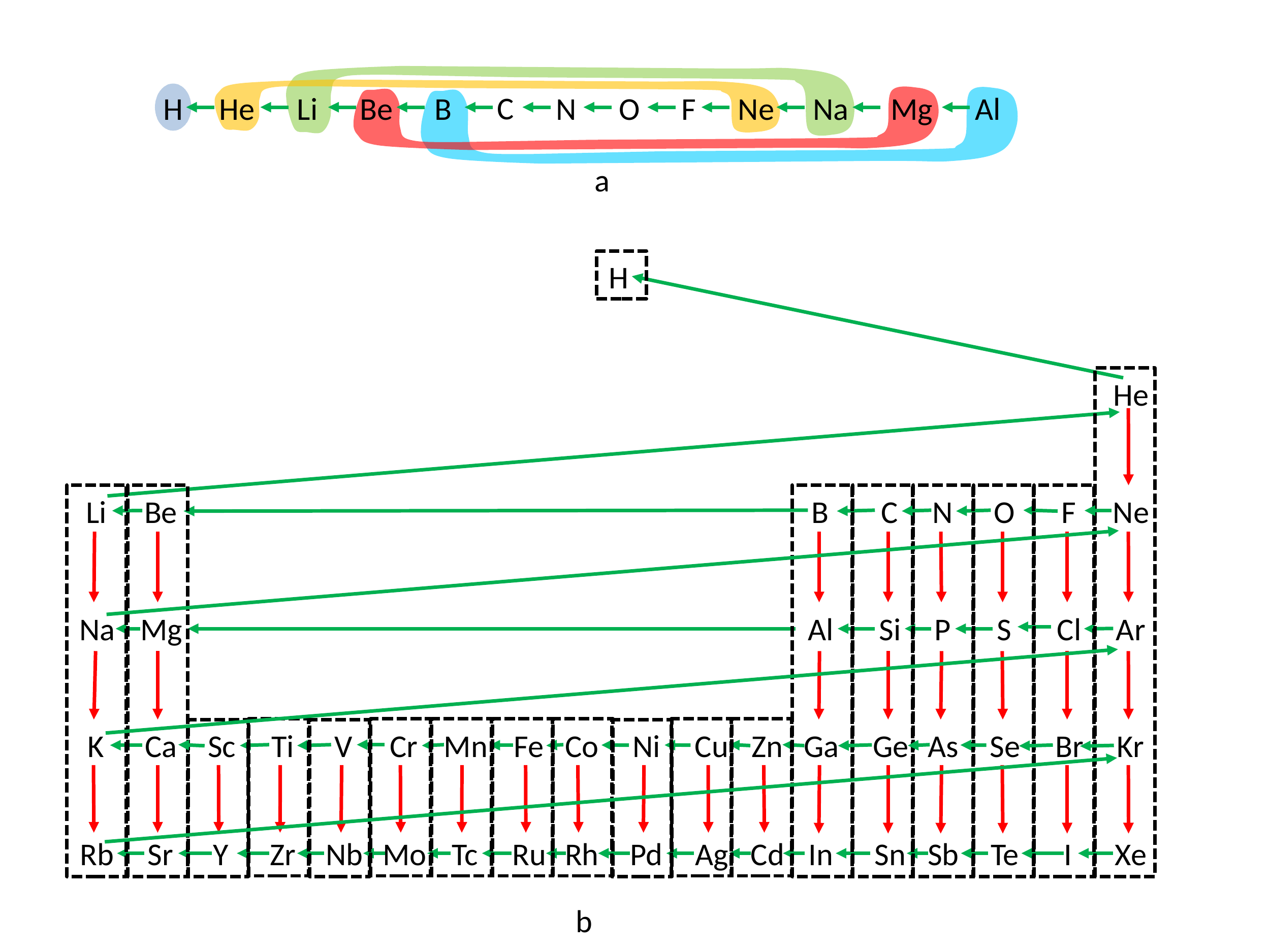}
	\caption{a) Order of elements, from H to Al, by $Z$ ($x\leftarrow y$ represents $Z(x)\preceq Z(y)$), where classes of similar elements are highlighted.  The bringing together of these classes, preserving order, leads to b) A periodic table, where red arrows indicate order relationships for elements inside similarity classes, whereas green arrows between elements of different similarity classes.}
	\label{PT-chem-elts}
\end{figure}

Once elements are ordered, it is the bringing together of similar elements that ``twist'' the order giving place to ``periods''.  Thus, the order by atomic number $Z$(Li) $<$ $Z$(Be) $< \ldots <$ $Z$(Ar) is twisted as $Z$(Li) $<$ $Z$(Be) $< \ldots <$ $Z$(Ne) and $Z$(Na) $<$ $Z$(Mg) $< \ldots <$ $Z$(Ar), for Li is brought together with Na, as they belong in a class.  The twist is also caused by the other classes: Be-Mg, B-Al, etc.  (Figure \ref{PT-chem-elts}a).  An important consequence of the twists is the oscillating behaviour of some properties of chemical elements, which are nothing else than the product of considering similarity classes and order simultaneously.

A structure capturing the aspects of Mendeleev's periodic system is given by the following:

\begin{definition}
Let $E$ be the set of chemical elements, $Z$ the atomic number, $\preceq_Z$ the order relation by $Z$, $P$ some properties of the elements, $C_P$ a classification by $P$; then the \emph{Mendeleevian periodic system} is the ordered partition  $(E, \preceq_Z, C_P)$.
\label{MendeleevianPS}
\end{definition}

A periodic table of the system highlighting the order relation $\preceq_Z$ among elements and among elements within classes is shown in Figure \ref{PT-chem-elts}b.

Similarity and order, however, can be treated in their broadest mathematical sense, giving place to richer structures, as we show in the next section.

\subsection{Generalised periodic system}

\begin{figure}[h]
	\centering
	\includegraphics[width=.7\textwidth,height=!, keepaspectratio]{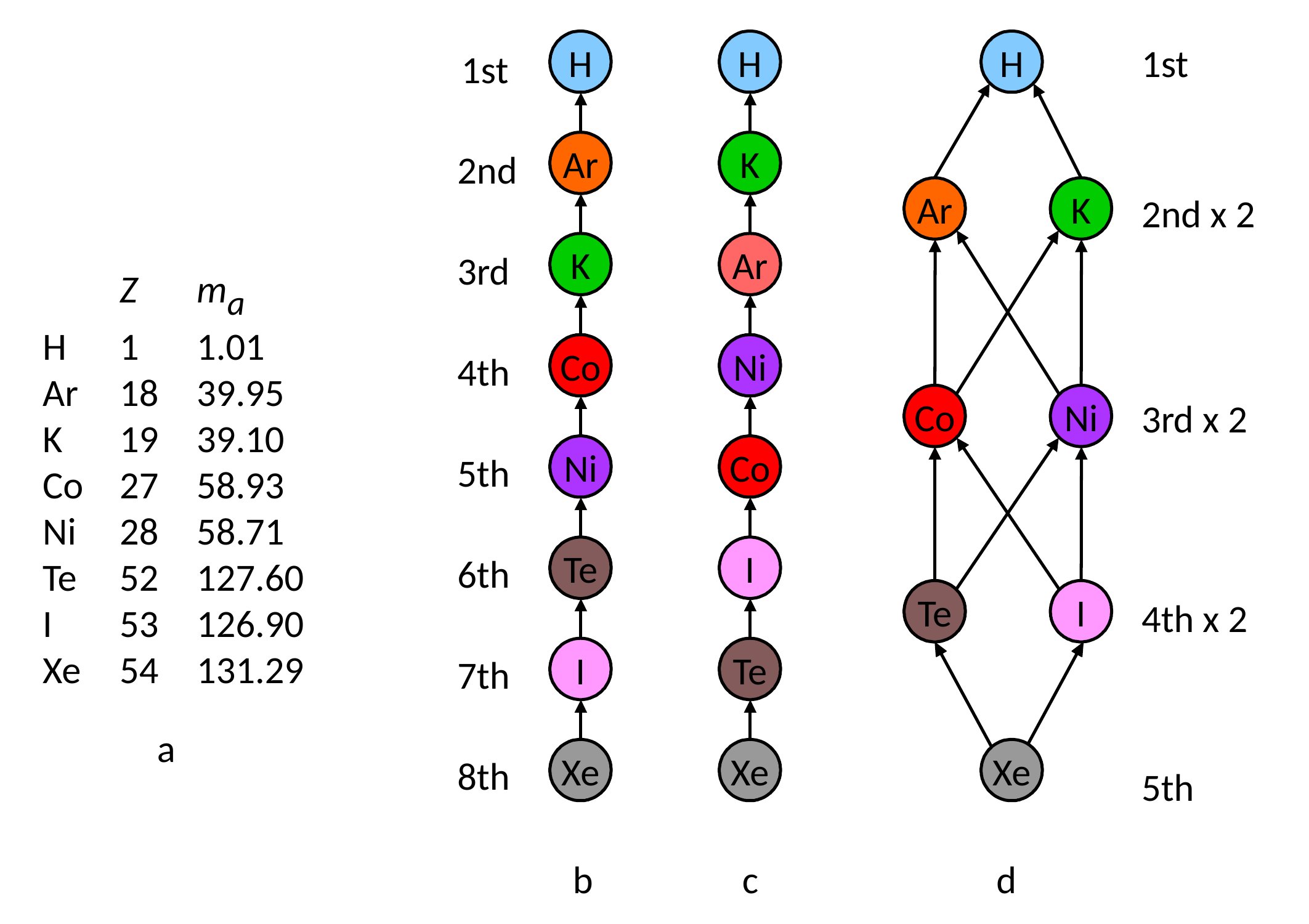}
	\caption{a) Some elements and their atomic numbers ($Z$) and atomic masses ($m_a$). b) Elements ordered by $Z$ and by c) $m_a$. d) Resulting order by simultaneously considering $Z$ and $m_a$.  In these Hasse diagrams (b to d) any sequence of arrows indicates that its extremes are comparable, e.g. in b, from the sequence H $\leftarrow$ Ar $\leftarrow$ K it follows that H $\preceq$ K; whereas the absence of such a sequence indicates that they are incomparable, e.g. Ar and K in d.}
	\label{HD-elts}
\end{figure}

Just as similarity is customarily based on more than one property, order can also be based on several properties \cite{rainerbook,EST1}.  Figure \ref{HD-elts} illustrates how eight chemical elements\footnote{They are selected to highlight the historical inversions of order resulting from atomic weight and atomic number.} can be ordered by atomic number $Z$ or by atomic weight $m_a$ (Figure \ref{HD-elts}a), independently (as usual), leading to Figures \ref{HD-elts}b and c, respectively.  Note that, when using either $Z$ or $m_a$, it is always possible to compare any pair of elements $x$ and $y$ and assess whether $x \preceq y$ or $y \preceq x$; in both cases it is said that $x$ and $y$ are \emph{comparable}. A set endowed with an order satisfying this property is called a \emph{total order}. However, when both $Z$ and $m_a$ are simultaneously used, conflicts among properties may arise, e.g. $Z$(Ar) $ < Z$(K) and $m_a$(Ar)  $>m_a$(K) (Appendix-Definition \ref{HDT-order}).  Therefore, it is no longer possible to claim that $x \preceq y$ or $y \preceq x$, in this case we say that $x$ and $y$ are \emph{incomparable}. An ordering allowing comparabilities and incomparabilities is called a \emph{partially ordered set} (Appendix-Definition \ref{order}). Figures \ref{HD-elts}b to d are graphical representations of partially ordered sets, called \emph{Hasse diagrams}, where an arrow $x \leftarrow y$ between two elements $x$ and $y$ is depicted only if $x \preceq y$ and there is no $z$ such that $x \preceq z \preceq y$.  This particular case of comparability is called a \emph{cover relation} and is denoted by $x \preceq: y$\footnote{Note that a cover relation is represented by adding a colon to the comparability.} (Appendix-Definition \ref{cover-preserving}).  Hence, in a Hasse diagram any comparability $x \preceq y$ is represented as a sequence of cover relations; for example, as comparability $Z$(H) $\preceq Z$(K), in Figure \ref{HD-elts}b, can be inferred from the cover relations $Z$(H) $\preceq: Z$(Ar) along with $Z$(Ar) $\preceq: Z$(K), therefore $Z$(H) $\preceq Z$(K) is not graphically represented. Likewise H $\preceq$ Ni, in Figure \ref{HD-elts}d, is inferred from H $\preceq:$ Ar and Ar $\preceq:$ Ni, or H $\preceq:$ K and K $\preceq:$ Ni\footnote{Note that $x\leftarrow y$ can also be the sequence, which occurs if and only if $x\preceq: y$.  Another depiction of Hasse diagrams takes the convention of replacing arrows by lines, where the direction of the order relation is inferred from the position of the elements on the drawing plane \cite{Trotter}.} (Appendix-Definition \ref{hasse-diagram}).

Thus, the ordering by more than one property may bring a structure of comparabilities and incomparabilities, which frees the periodic system from the historical total order imposed by $Z$.

A further generalisation of periodic system can be obtained if classification is analysed.  Here the question that arises is whether a classification leading to partitions is meaningful and general enough for the system of chemical elements.  Is it always desirable to have disjoint classes?  Could partitions be instances of a more general case for periodic systems?  Chemistry helps to solve these questions.  It has been found that a chemical element may belong to more than one class of similar elements,\footnote{Perhaps this is also the case of the discussion about which elements belong in group three of the periodic system of chemical elements \cite{group3}.} as is the case of Ti and Mn \cite{Rayner-Canham2018}.  Other studies of hierarchies of similarity classes show that elements belong to multiple classes with different degrees of similarity \cite{RestrepoACS}, which contrast with the rigid structure of partitions of similarity classes, proper of the Mendeleevian system \cite{Rouvray}.  Therefore, a more general structure for a periodic system is a collection of subsets of similar elements endowed with an order relation.

\begin{figure}[h]
	\centering
	\includegraphics[width=.4\textwidth,height=!, keepaspectratio]{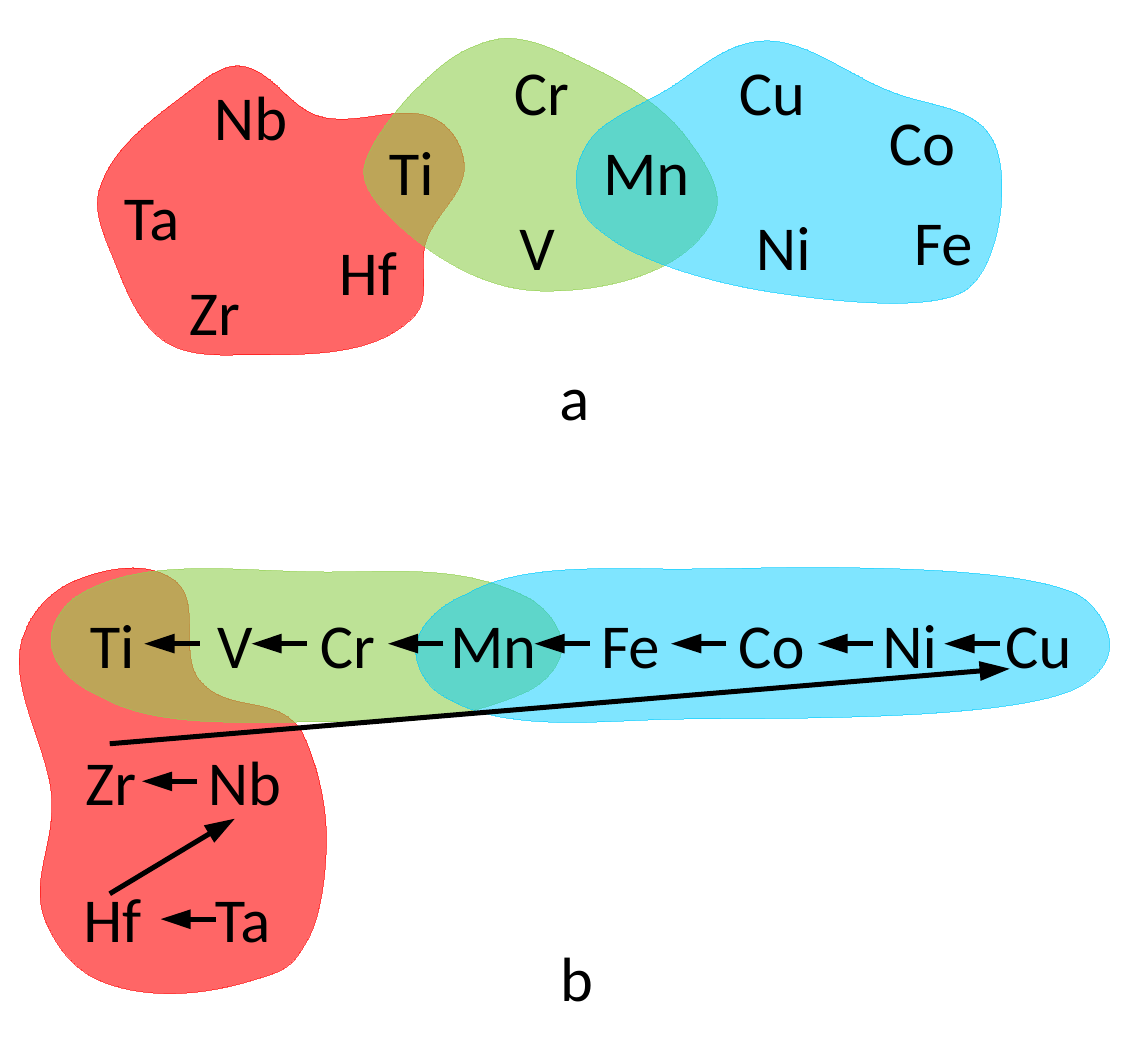}
	\caption{a) Hypergraph with three hyperedges (similarity subsets) and b) ordered hypergraph, where the order relation is given by the arrows.}
	\label{Ti-Mn}
\end{figure}

A mathematical object made of collections of subsets, called \emph{hyperedges}, is that of \emph{hypergraphs} (Appendix-Definition \ref{hypergraph}) \cite{Berge}.  Elements belonging to a hyperedge (subset) are regarded as related, and the nature of their relation depends on the system to be modelled \cite{Klamt}, in our case, the relation is similarity. Figure \ref{Ti-Mn}a shows three similarity subsets with two overlaps caused by the dual similarities of Ti and Mn \cite{Rayner-Canham2018}.  This system corresponds to a hypergraph.

However, the periodic system is not complete without ordering.  Therefore, the following definition provides a general structure for the periodic system:

\begin{definition}
Let $E$ be a non-empty set, $A$ a set of properties, $\preceq_A$ the order relation by $A$, $P$ some properties of the elements in $E$, $C_P$ a collection of subsets of similar elements regarding $P$ and $H=(E,C_P)$ a hypergraph; then the ordered hypergraph\footnote{Note that several authors \cite{Bollobas, Eslahchi} refer to ordered hypergraphs as $(H,<)$, with $H=(E,C_P)$; where $<$ is a linear order of the elements of $E$, i.e. the elements belonging in a subset (hyperedge) are totally ordered.  In contrast, our definition of ordered hypergraph is more general by allowing general orders $\preceq$, which corresponds to the definition of partial-order hypergraph in \cite{He}.} $(H,\preceq_A)$ is a \emph{periodic system}.
\label{defps}
\end{definition}

\begin{figure}[h]
	\centering
	\includegraphics[width=.4\textwidth,height=!, keepaspectratio]{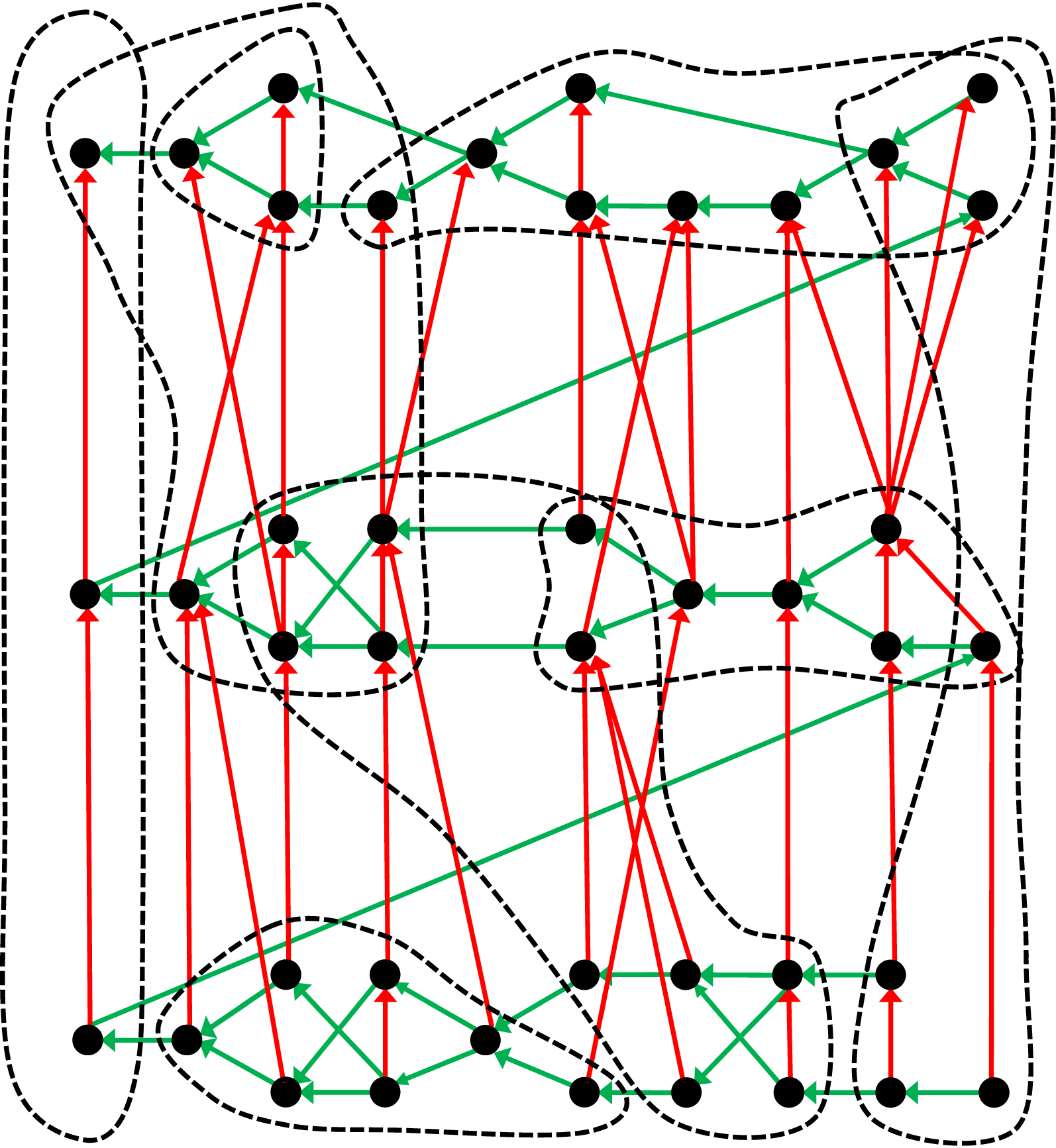}
	\caption{A periodic system (ordered hypergraph) where elements $E$ are nodes, which are ordered by $\preceq_A$ (green arrows) and whose order within similarity subsets (hyperedges represented as dotted lines) is shown as red arrows.}
	\label{PT-abstract}
\end{figure}

The periodic system corresponding to Figure \ref{Ti-Mn}a is shown in Figure \ref{Ti-Mn}b, where the system of similarity subsets (Figure \ref{Ti-Mn}a) is endowed with an order (arrows), in this case given by atomic number.  A depiction of a general periodic system is shown in Figure \ref{PT-abstract}, where the partially ordered structure and the generality of the collection of subsets is highlighted.
 
Definition \ref{defps} shows that the Mendeleevian periodic system is one of the possible periodic systems.  It comes up by ordering by atomic number, an order that has not incomparabilities (total order), and by taking subsets of similar elements, which depends on the properties used for the classification\footnote{In reference \cite{Jensen} (paper 2), Mendeleev discusses different classification criteria and also several for ordering.} and that leads to a partition, as a particular case of a collection of subsets.  This chemical freedom in classification, which contrasts with the conservative ordering by $Z$, is the cause of the several periodic systems of chemical elements, which when combined with their possible representations gives place to the more than thousand periodic tables of the Mendeleevian periodic system.

Definition \ref{defps} frees the periodic system from the chemical domain, for it can now be used in other contexts as long as the elements be provided, as well as the criteria for ordering and classifying them.  However, not to go far from chemistry, we show in the next section how Definition \ref{defps} can be used to devise a periodic system of bonds, in contrast to the traditional one of chemical elements.

\section{A periodic system of polarization of single covalent bonds}

Polarizability, i.e. the tendency of charge distribution to be distorted in response to an external electrical field, is an important property of materials at different levels, ranging from atomic and molecular to bulk scales \cite{Rupasinghe2015}.  Its importance is given by its relationship with, e.g. stiffness of materials, compressibility and other properties \cite{Rupasinghe2015}.  Not to mention its pedagogical chemical value \cite{JCE}.  By addressing polarizability at a simple molecular level of atoms forming single covalent bonds, here we devise a periodic system tailored to such bonds.

Note that bond polarization is based on the definition of atomic charge, of which there are several, from different theoretical and experimental perspectives \cite{Schwarz1994}.  Moreover, there are different properties to characterise bond polarizability, e.g. electronegativity, atomic radius, ionization potential, electron affinity, atomic volume and some others coming from natural bond orbital treatments, among others.  In any case, the characterisation requires at least two properties related to the potential nucleus-electron attraction and the kinetic repulsion of electrons that make a single covalent bond a stable system.  Two reasonable properties meeting this condition and readily available are electronegativity \cite{Pauling} and bond distance, as expressed by atomic radius of bonded atoms\footnote{Bond characterisation may also be attained through specific or averaged properties.  Properties selected in this paper are averaged, but specific ones such as Allred-Rochow electronegativities and radii with reference to a particular parent group, e.g. methyl, could also be used.  A systematic approach for the selection of properties, given a response variable, has been recently published in \cite{PhysRevMaterials.2.083802}.} \cite{Pyykko2009}.

\begin{figure}[h!]
\centering
	\includegraphics[width=.9\textwidth,height=!, keepaspectratio]{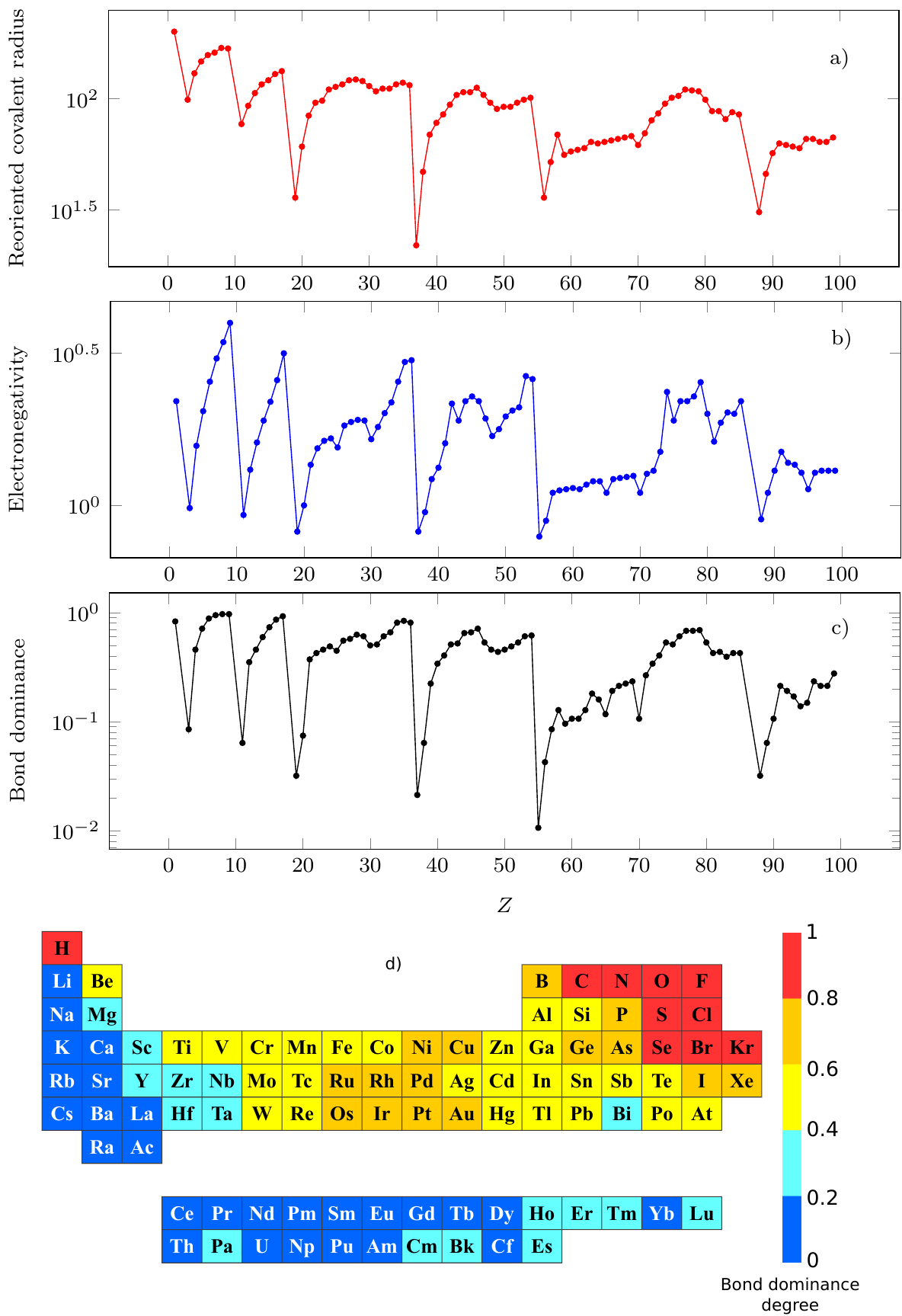}
	\caption{Logarithmic distribution of a) reoriented single-bond covalent radius (pm) and b) Pauling electronegativity (dimensionless) and of c) bond dominance degree (dimensionless) over the elements ordered by atomic number.  d) Conventional periodic table with elements classified by their bond dominance degree.  Elements in the red hyperedge correspond to those forming the most polarized bonds, while those in the dark blue to the least.}
	\label{3plots}
\end{figure}

We considered 94 single covalent bonds ($E$) of the form $x-y$, where $y$ is a chemical species as explained latter and $x$ is a chemical element.\footnote{The elements $x$ are: H, Li, Be, B, C, N, O, F, Na, Mg, Al, Si, P, S, Cl, K, Ca, Sc, Ti, V, Cr, Mn, Fe, Co, Ni, Cu, Zn, Ga, Ge, As, Se, Br, Kr, Rb, Sr, Y, Zr, Nb, Mo, Tc, Ru, Rh, Pd, Ag, Cd, In, Sn, Sb, Te, I, Xe, Cs, Ba, La, Ce, Pr, Nd, Pm, Sm, Eu, Gd, Tb, Dy, Ho, Er, Tm, Yb, Lu, Hf, Ta, W, Re, Os, Ir, Pt, Au, Hg, Tl, Pb, Bi, Po, At, Ra, Ac, Th, Pa, U, Np, Pu, Am, Cm, Bk, Cf, Es; which are the elements considered in \cite{Leal} and shown in Figure \ref{3plots}.}  The properties for ordering are $A=$\{Pauling electronegativity, single-bond additive covalent radius\}.  By additive covalent radius of bond $x-y$ is meant $r(x-y)=r(x)+r(y)$.  These radii were obtained from either experimental or theoretical data of chemical species including bonds of the sort $x-x$, $x-\text{H}$, $x-\text{CH}_3$ and $x-y$, being $x$ and $y$ different chemical elements \cite{Pyykko2009}.  The similarity  subsets (hyperedges) $C_P$ for the bonds $x-y$ were based on the resemblance of the elements $x$ when forming binary compounds, which yielded 44 classes \cite{Leal} (Table \ref{44classes}).  These classes are based on the chemical idea that elements are similar if they combine with the same substances to produce chemically similar compounds \cite{Schummer1998}.  For instance, alkali metals are similar because they combine with water to produce alkalies, which when combined with hydracids of halogens produce simple salts, e.g. LiF, NaF, LiCl and NaCl, etc.  Hence, alkali metals are regarded as similar, for a large amount of the compounds they form are common to the similarity class of alkali metals.  As the notion of common compound is central to describe similarity of chemical elements, it was formalised in \cite{Leal} for binary compounds as follows: two elements $x$ and $y$ have a common binary compound if there exists a third element $z$, and binary compounds $x_az_b$ and $y_az_b$.  Therefore, similarity between $x$ and $y$ increases with the number of common binary compounds they have.  In \cite{Leal}, the notion of common compound includes similarity in proportions of combination to differentiate, for instance, between alkali and alkali earth metals.  In such a setting, LiF and BeF$_2$ are not common binary compounds of Li and Be, but BeF$_2$ and MgF$_2$ are for Be and Mg.  Having a set of binary compounds, similarity between two elements $x$ and $y$ is calculated as the number of common compounds between them, leading to a similarity matrix upon which classification algorithms may be applied, e.g. cluster analysis, yielding similarity classes \cite{Leal}.

The set of properties $P$ for the classification of single covalent bonds is made by the global neighbourhoods of each element.  All calculations on order relationships here reported were performed with the free-ware Python package PyHasse \cite{Voigt2010,Rainer2013} developed by Rainer Bruggemann.\footnote{More information on PyHasse can be requested to its developer at \href{mailto:brg\_home@web.de}{brg\_home@web.de}.}  This package has an on-line version \cite{PyHasse}, where some further order calculations are possible.

\begin{table}[!h]
\centering
\caption{44 similarity classes of chemical elements.}
\label{44classes}
\begin{tabular}{ c | c | c| c}
 H & B & C & N \\ 
    O & Si & P & S \\
    Ti & Cr & Mn & Au \\
    Bi & Po & La & Ce \\ 
    Ac & Pu & Ir,Rh & Se,Te \\ 
    Y,Sc & Ca,Mg & Mo,W & Ag,Cu \\ 
    As,Sb & Am,Cf & Tb,Pr & Li,Na \\ 
    Cm,Bk,Es & Sm,Eu,Yb & Be,Sr,Ba & K,Rb,Cs \\ 
    V,Nb,Ta & Zn,Cd,Hg & Ru,Os,Pt & Ge,Sn,Pb \\ 
    F,Cl,Br,I & Kr,Xe,At,Ra & Tm,Dy,Pm,Nd & Lu,Er,Ho,Gd \\ 
    Tc,Re,Pa,Np & Fe,Co,Ni,Pd & Al,Ga,In,Tl & Zr,Hf,Th,U \\

\end{tabular}
\vspace*{-4pt}
\end{table}

\begin{figure}[h]
	\centering
	\includegraphics[width=.6\textwidth,height=!, keepaspectratio]{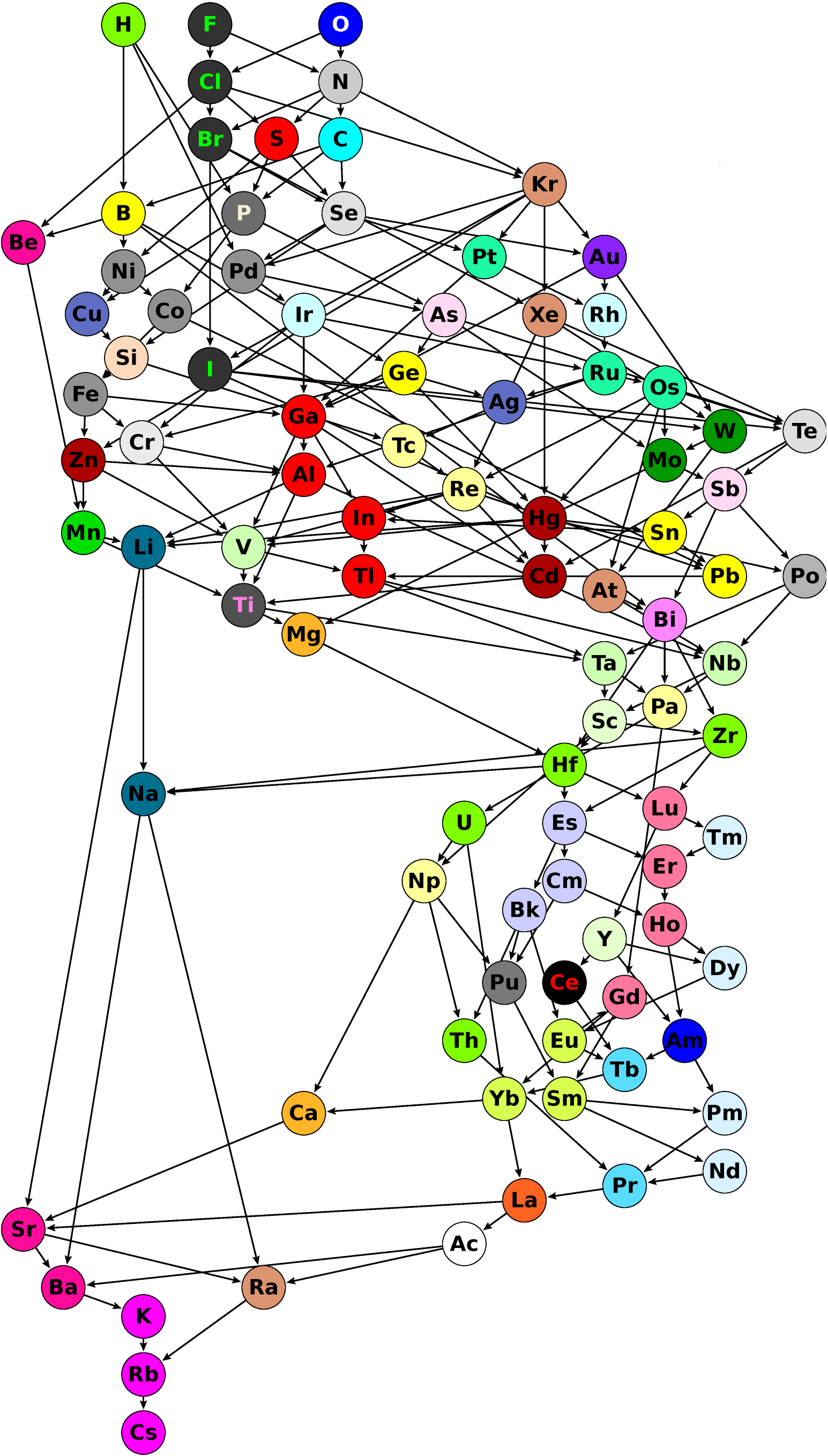}
	\caption{A periodic system of polarized single covalent bonds, where 94 bonds $x-R=b(x)$, with $x$ a chemical element and $R$ another chemical species, are ordered by Pauling's electronegativity of element $x$ and its reoriented atomic radius in single bonds.  For simplicity $b(x)$ is labelled as $x$.  There are 93 nodes, for Bk and Cf are equivalent and are represented by Bk.  At the top of the system highly polarized bonds appear and at the bottom the least ones.  Hyperedges (subsets of similar elements as shown in Table \ref{44classes}) are depicted as bonds sharing filling and font colours.}
	\label{PT-radii-electronegativity}
\end{figure}

To attain an ordering with chemical meaning where more polarized bonds involve electronegative elements and hold short radii, we reoriented the radius of element $x$, as $\bar r(x)=\text{max }r-r(x)$.  In Figure \ref{3plots} it is seen that electronegativity and reoriented radius are highly correlated, as quantified by the 0.83 Spearman correlation.\footnote{Spearman coefficient quantifies whether two variables are monotonically related, not necessarily linearly, ranging from 0 (not correlated) to 1 (correlated).}  However, these two properties lead to several incomparabilities in the periodic system, as we show later.  A correlation of 1 would indicate that electronegativity and oriented radius hold the same total order of bonds, i.e. that when ordered by $\bar r(x)$ the order obtained is the same than when bonds are ordered by electronegativity of $x$.  As the correlation is high, a large number of comparabilities is expected (few incomparabilities).  On the other hand, since bonds involving Bk and Cf are equivalent, for they have the same electronegativity and $\bar r$, there are 93 different representative bonds accounting for $93\times 92/2=4,278$ order relationships, which are split in 3,548 comparabilities and 730 incomparabilities, i.e. 83\% of the order relationships are comparabilities and 17\% are incomparabilities.  A depiction of this periodic system is shown in Figure \ref{PT-radii-electronegativity}.\footnote{As data uncertainty may affect the ordering, a fuzzy set theoretical approach has been devised to analyse uncertainty effects \cite{Rainer2011, Rainer2012}, where the order relation $\preceq$ is replaced by a fuzzy inclusion relation.  In such a setting the ordering of any two elements is not any more given by the pairwise comparison of their properties (Appendix-Definition \ref{HDT-order}) but by the degree of subsethood of the two elements, now considered as sets of their property values.}

The most polarized bonds correspond to those involving H, F and O (at the top of Figure \ref{PT-radii-electronegativity}), and the least is Cs (at the bottom).  To know if a bond $b(x)=x-R$, with $x$ an atom of element $x$ and $R$ another chemical species \cite{Pyykko2009}, is more polarized than another $b(y)$, i.e. if $y-R \preceq x-R$, it must be found in Figure \ref{PT-radii-electronegativity} a sequence of arrows from $b(x)$ to $b(y)$\footnote{In mathematical terms it corresponds to finding a chain between $x$ and $y$ (Appendix-Definition \ref{chain}).} \cite{Trotter}; this is the case of, e.g. $b($I$) \preceq$ $b($F$)$.

A consequence of exploring order relationships is seen, e.g. in the case of bonds of Cl and N, or Pu and Am, which are incomparable, i.e. with no sequence of arrows connecting them.

Being located at the top of the periodic system does not necessarily mean that such bonds are more polarized than the others, e.g. $b($H$)$ is more polarized than 77 other bonds, while $b($O$)$ and $b($F$)$ more than 90 bonds; $b($H$)$, $b($O$)$ and $b($F$)$ are all at the top of the system.  To quantify this degree, we devised the bond dominance degree:

\begin{definition}
Let $x$ and $y$ be single covalent bonds $b(x)$ and $b(y)$, with $x\neq y$.  The \emph{bond dominance degree} of $x$ is given by:
\[Dom(x):=\frac{C_{y \prec x}}{n-1}\]
where $C_{y \prec x}=\vert\{y : y\prec x\}\vert$ and $n$ is the number of bonds. Note that $y \prec x$ indicates those $y$ such that $y \preceq x$ but that are not $y = x$.\footnote{Hence, $\prec$ is a relation that is not reflexive, it only holds antisymmetry and transitivity, while an order relation $\preceq$ holds the three of them.}
\label{eldomdeg}
\end{definition}

$Dom(x)=1$ indicates that all other bonds different from $x$ are dominated by $x$, i.e. that $b(x)$ is more polarized than all the other bonds.  $Dom(x)=0$ shows that $x$ is less polarized than any other bond.  A plot depicting the bond dominance degree is shown in Figure \ref{3plots}c, where, keeping the chemical tradition, bonds are ordered by the respective $Z$ of the bonded atom $x$.  Figure \ref{3plots}b is actually a plot of the function $(Z,f(Z))$, where $f(Z)=Dom(Z)$, which is an oscillating function resulting from the oscillating nature of electronegativity and single-bond covalent radius (Figures \ref{3plots}a and b).

Figure \ref{3plots}d depicts the conventional periodic table with elements coloured by their bond dominance degree.  Therein alkali metal, heavy alkaline earth and most of the lanthanoid and actinoid bonds are more polarized than only 20\% of the other bonds.  Mg bond is more polarized than 40\% of the others, as some early transition metal bonds, e.g. Sc, Y, Zr and Hf.  Be bond, whose Be is similar to Sr and Ba (Table \ref{44classes}), is more polarized than the bonds of Sr and Ba.  In fact, $b($Be$)$ is more polarized than 60\% of the other bonds, as several transition metal and non metal bonds.  Most of the platinum and coinage metals (except Ag), as well as B, P, Ge, As, I and Xe (dark yellow), form more polarized bonds than 80\% of the other bonds.
%Most of the platinum and coinage metals, except Ag, form more polarized bonds than 80\% of the other bonds, as B, P, Ge, As, I and Xe.
The bonds that are more polarized than the rest of the bonds are those of H, C, N, Kr, halogens and chalcogens, except Te.  Specific details on the dominated, dominating and incomparable bonds for each bond\footnote{Bond $y$ is a dominating bond of $x$ if the relation $x \prec y$ holds.  In mathematical terms the dominated bonds of $x$ correspond to the \emph{down set} or \emph{ideal} of $x$ without the bond $x$.  Likewise, the dominating bonds of $x$ are the \emph{up set} or \emph{filter} of $x$ without the bond $x$ \cite{Trotter}.} are provided in Table S1.

So far, it has been discussed how order and classification shed light on bond polarization.  However, the two relations can be further considered to explore order relations within and among subsets of bonds.  These relationships correspond, respectively, to the red and green arrows in Figures \ref{PT-chem-elts} and \ref{PT-abstract}.

\subsection{Ordering bond polarizations within subsets of similar elements (hyperedges)}
The analysis of order relationships within hyperedges allows assessing whether a subset of bonds of similar elements also involves an ordered structure.  This order relationship permits knowing whether, e.g. the well-known electronegative fluorine forms most polarized bonds than the other halogens.  To analyse these within-hyperedge order relations, we quantified the degree of within-hyperedge comparability.

\begin{definition}
Let $C$ be a subset of single covalent bonds (hyperedge), the \emph{within-hyperedge dominance degree} $Dom(C)$ is given by
\[
Dom(C):=\frac{2T_{j\prec i}}{n(n-1)}
\]
with $n$ being the number of bonds in $C$ and $T_{j\prec i}=\vert\{(x_i,x_j): x_j\prec x_i, x_i,x_j \in C\}\vert$.
\label{withindegree}
\end{definition}

Hence, for a hyperedge with $n$ bonds, $n(n-1)/2$ relationships of the sort $x \prec y$ are expected, with $x,y \in C$.  How many of them are actually $\prec$ (non self-comparabilities) is what within-hyperedge dominance degree quantifies.  Note that this degree is only calculated for hyperedges of more than one bond, for the relation $\prec$ does not allow self comparisons.  Hyperedges where all relationships are comparabilities, i.e. where there is a chain (Appendix-Definition \ref{chain}) containing the bonds of the hyperedge, are robust in terms of similarity and order.  These hyperedges have 1 as degree of within-hyperedge comparability; likewise, if the bonds of the hyperedge are not comparable at all, the degree of comparability is 0.

\begin{figure}[h]
	\centering
	\includegraphics[width=.8\textwidth,height=!, keepaspectratio]{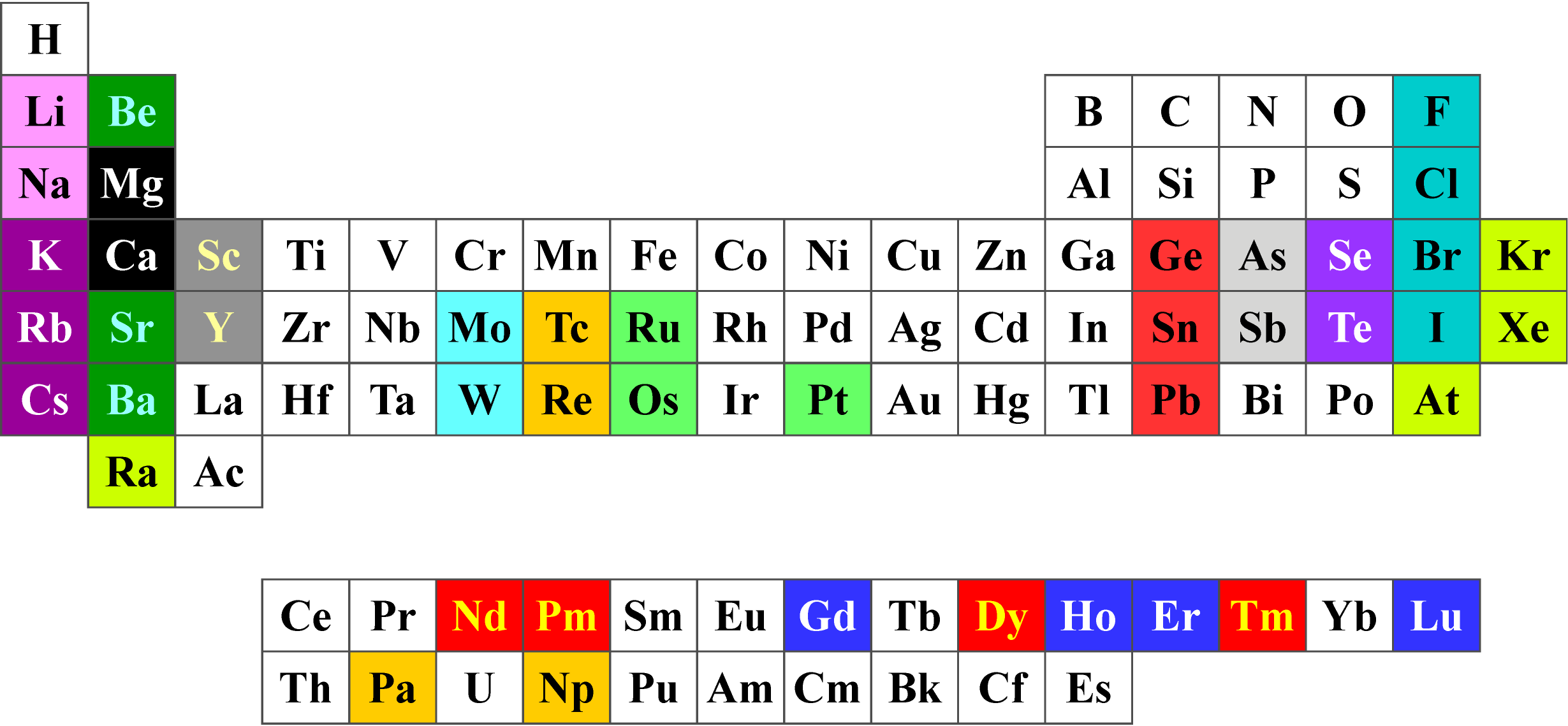}
	\caption{Periodic table where hyperedges with within-dominance degree equal to 1 are coloured.  Elements (bonds) sharing cell colour and font belong in the same hyperedge and hold dominance degree 1.}
	\label{PT-dom1}
\end{figure}

There are 26 non-single hyperedges of similar chemical elements, out of the 44 discussed (Table \ref{44classes}), whose degrees of comparability are shown in (Table \ref{deg-comp-incomp-classes}).  It is seen that almost half of the hyperedges have within-hyperedge comparability degrees greater than 0.5.  This shows that these hyperedges not only gather bonds of similar elements, but that they have a rich order structure.  This makes that similar elements, e.g. Ge, Sn and Pb, that form a hyperedge with $Dom(C)=1$, can be ordered by bond polarization, in this case being $b($Ge$)\succ b($Sn$)\succ b($Pb$)$.  This kind of trend is well-known for halogens, which actually form a hyperedge with $Dom(C)=1$, being $b($F$)$ the most polarized single covalent bond.  There are hyperedges with non-vertical similarities on the table having $Dom(C)=1$, e.g. $b($Tc$)\succ b($Re$)\succ b($Pa$)\succ b($Np$)$ and $b($Ru$)\succ b($Os$)\succ b($Pt$)$.  Figure \ref{PT-dom1} shows the hyperedges with $Dom(C)=1$.  There is an average of 0.73 for within-hyperedge dominance degree, which is expected given the high amount of comparabilities in the periodic system.

\begin{table}[!h]
\centering
\caption{Within-hyperedge dominance degree for the 26 non-single subsets of similar bonds.  Hyperedges with degree 1 are shown in non-increasing order of polarizability, e.g. Ge $\succ$ Sn $\succ$ Pb.  For simplicity $b(x)$ is labelled $x$.}
\label{deg-comp-incomp-classes}
\begin{tabular}{  l | c }

    Subset (Hyperedge) & \makecell{Within-hyperedge \\dominance degree} \\ \hline
    Ge, Sn, Pb & 1  \\ 
    Zr, Hf, Th, U & 0.5  \\ 
    Al, Ga, In, Tl & 0.66  \\ 
    Am, Cf & 0  \\ 
    As, Sb & 1  \\ 
    Cu, Ag & 0  \\ 
    Ru, Os, Pt & 1  \\
    Mo, W & 1  \\
    Fe, Co, Ni, Pd & 0.5  \\ 
    Zn, Cd, Hg & 0.33  \\ 
    V, Nb, Ta & 0.66  \\ 
    Tc, Re, Pa, Np & 1  \\ 
    Li, Na & 1  \\ 
    K, Rb, Cs & 1  \\ 
    Ca, Mg & 1  \\ 
    Be, Sr, Ba & 1  \\ 
    Lu, Er, Ho, Gd & 1  \\ 
    Tm, Dy, Pm, Nd & 1  \\ 
    Sm, Eu, Yb & 0.66  \\ 
    Y, Sc & 1  \\ 
    Cm, Bk, Es & 0.66  \\ 
    Tb, Pr & 0  \\ 
    Ra, Kr, Xe, At & 1  \\ 
    Ir, Rh & 0  \\ 
    Se, Te & 1  \\ 
    F, Cl, Br, I & 1 \\

\end{tabular}
\vspace*{-4pt}
\end{table}

\subsection{Ordering bond polarizations among subsets of similar elements (hyperedges)}
By ordering hyperedges we can address questions like: Are lanthanoid bonds more polarized than actinoid ones?, which have technological, as well as geochemical implications related to the materials they may form and the extraction from ores these elements undergo.  Similar questions aiming at comparing sets of elements can be addressed.  To do so, we applied the dominance degree for hyperedges.

\begin{definition}
Given $C_i$ and $C_j$ as hyperedges of bonds, the \emph{inter-hyperedge dominance degree} $Dom(C_i,C_j)$ of $C_i$ over $C_j$ is given by:
\[Dom(C_i,C_j):=\frac{T_{j\prec i}}{n_in_j}\]
with $T_{j\prec i}=\vert\{(x_i,x_j) : x_i\in C_i, x_j\in C_j, x_j\prec x_i\}\vert$.
\label{domdegclass}
\end{definition}

\begin{figure}[h]
	\centering
	\includegraphics[width=\textwidth,height=!, keepaspectratio]{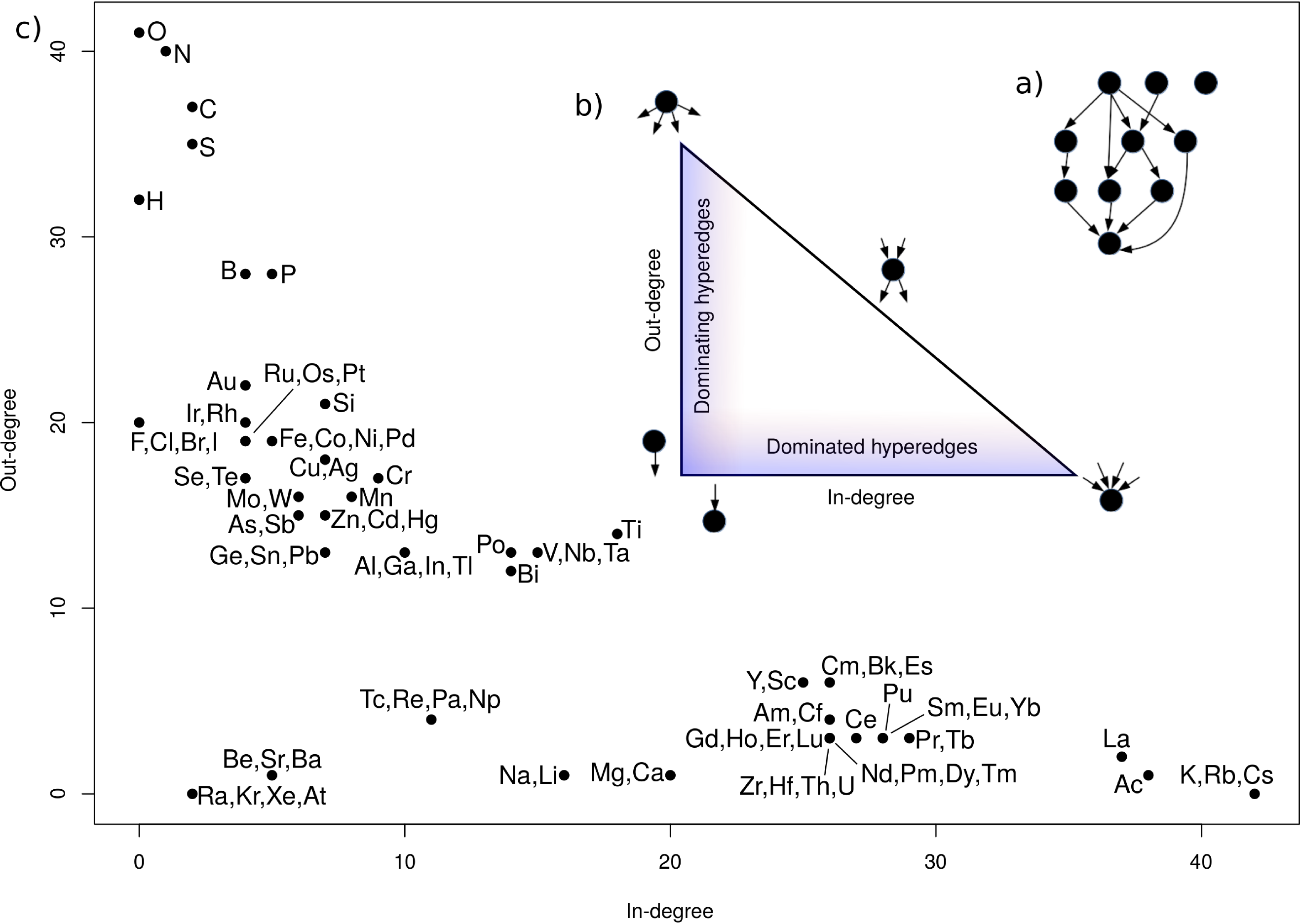}
	\caption{a) Dominance diagram of 10 hyperedges (black nodes) and its respective b) dominance profile, where some hyperedges are highlighted.  The diagonal limiting the profile corresponds to all possible sums of in and out-degrees being equal to the number of hyperedges.  c) Dominance profile for the 44 hyperedges of single covalent bonds.}
	\label{in-out-deg-scatter}
\end{figure}

Hence, for a given couple of hyperedges of bonds $C_i$ and $C_j$, $Dom(C_i,C_j)$ quantifies how many bonds of $C_i$ dominate those of $C_j$, i.e. how many bonds of $C_i$ are more polarized than bonds in $C_j$.  Figure \ref{in-out-deg-scatter}a shows a schematic representation of a \emph{dominance diagram} \cite{EST1}, where the most dominated hyperedges are at the bottom with a high number of incoming arrows (high in-degree) and most dominating hyperedges are located at the top, holding high number of outgoing arrows (out-degree).\footnote{Note that the arrows of the dominance diagram are not cover relations as in a Hasse diagram.  In particular, dominance diagrams do not hold transitivity \cite{EST1, Restrepo2008}.} When the dominance diagram turns too complex (with many arrows), it is better to depict each hyperedge in a coordinate system given by its in- and out-degrees (Figure \ref{in-out-deg-scatter}b), which we call the \emph{dominance profile}.\footnote{Similar diagrams to represent complex partially ordered sets are devised in \cite{Rainer2013,Quintero2018}.}  As the dominance diagram for hyperedges of bonds is too complex, even for $Dom(C_i,C_j)>0.95$; we show in Figure \ref{in-out-deg-scatter}c its respective dominance profile.

Figure \ref{in-out-deg-scatter}c shows that the least polarized bonds are those where the most electropositive alkali metals, La and Ac are involved.  These bonds are dominated by almost all other hyperedges and they dominate no other hyperedge or just a couple of them.  A cluster of a bit more polarized bonds is made by those involving some transition metals such as \{Y, Sc\},  electropositive alkaline earths \{Mg, Ca\} and most of the lanthanoids and actinoids.  Bonds with intermediate dominances are \{V, Nb, Ta\} and Ti that are more polarized than about one third of the other hyperedges and less polarized than about half of the other hyperedges.  A cluster of dominating bonds is made by several transition metals, with in-degrees about 8 and out-degrees close to 20.  These hyperedges of bonds of metals are more polarized than about half of the other hyperedges and are only less polarized than about 8 others, where O, N, C, S and H bonds are included.  It is found that actinoids dominate more hyperedges than lanthanoids, except for La and Ac, where La dominates more hyperedges than Ac.

\section{Conclusions and outlook}

Based on an analysis of the periodic system of chemical elements, we have formalised and generalised the periodic system as a set endowed with a system of similarity classes, whose elements hold an order relation. This structure corresponds to an ordered hypergraph, where similarity classes are hyperedges.

An advantage of having a mathematical structure for a periodic system is that it opens the possibility of exploring and formally characterising the relationships among periodic systems, i.e. given two periodic systems $P_1$ and $P_2$ , it can be determined whether one is a substructure of the other $P_1 \subseteq P_2$, if they are isomorphic $P_1 \simeq P_2$ , equivalent $P_1 \equiv P_2$ or equal $P_1 = P_2$ (Appendix-Definitions \ref{hypergraph} to \ref{hypergraph-isomorphism}). This brings up new questions. How many different periodic systems of the chemical elements have been devised? Which of them are isomorphic or equivalent? (Appendix-Definitions \ref{hypergraph-equivalence} and \ref{periodic-system-relations}).  Which systems are the most populated by their projections into periodic tables? Which is the super-structure formed by all the devised periodic systems? Are there some sort of embedding relations between them?

The structure here reported is flexible enough to accommodate new chemical elements, all of them located in the region of superheavy elements (SHEs), right after oganesson (Z=118).\footnote{Current estimations indicate that $Z$=173 is the heaviest possible element \cite{Indelicato2011}.}  Although the hypergraph structure was actually the framework in which Mendeleev predicted elements and several of their properties; he did it through interpolative methods.  This is no longer possible because the expansion of the system is in the SHE region.  Instead, predicting new elements requires relativistic quantum theoretical methods \cite{Pyykko2011}, which is how some SHEs properties are addressed, e.g. ionization potentials.  Once such calculations are provided, elements can be classified using properties derived from relativistic methods, e.g. electronic configurations in the ground state of the neutral atom.\footnote{As noted by Haba \cite{Haba2019}, the electronic configurations of SHEs are difficult to estimate, for valence orbitals are energetically close to each other.  This is especially difficult for elements with $Z >$121.} As SHEs have associated atomic numbers $Z$, these elements can therefore be ordered. Thus, new elements can be incorporated into the structure, for their similarity classes can be determined as well as their order relationships.

Although there is no room for interpolations, structures encoding chemical information about similarity and order can be used, as shown by Klein and coworkers \cite{Klein1995, Restrepo2011, Panda2013}.\footnote{A case in point is the ordering of substituted cubanes, which were ordered by the relation established between molecular structures when one can be obtained, by H-substitution, from the other \cite{Restrepo2011}.  By using three different interpolative methods that take into account the order relationships of couples of cubanes, not experimentally measured densities of nitro-cubanes could be estimated.  The results showed that the estimation of known densities were very close to experimental values.}  They have estimated properties of unknown substances, which makes foreseeable using the ordered hypergraph structure to estimate unknown properties of known elements.

Another instance of the relevance of the structure of the system is its recent use in the prediction of enthalpies of formation of several compounds \cite{C8SC02648C}.  There, Zhang and coworkers show how sensitive their neural networks predictions are to the input structure, which is a periodic table. As we have discussed, a periodic table is a mapping of the ordered hypergraph to a bi-dimensional space and there are many possibilities for the mapping.  The striking result is that by randomising the input structure, the quality of the estimations drops down. Also relevant is that the input table is a traditional one containing the most well-known similarity classes of chemical elements, ordered by atomic number, and that such a table yields the best predictions.  The destruction of such a structure by shuffling the elements reveals how important the ordered hypergraph is for the system.

Taking into account the structure of periodic system, we devised a periodic system for polarized single covalent bonds, which shows not only the similarity and order relationships for bonds, but allows exploring order relationships inside classes of similar bonds and among classes of bonds. This last order was of interest for Mendeleev, as noted when writing ``The objective of my memoir will be fully achieved if I can successfully direct the attention of investigators to those relationships involving the atomic weights of dissimilar elements, which, as far as I know, have so far been entirely ignored'' (p. 145 in \cite{Jensen}).

We found that most of the classes of similar bonds have an internal ordered structure, ranging from the typical example of halogens, where bonds of F are more polarized than those of Cl, Br and I; to cases involving transition metals and actinoids as \{Tc, Re, Pa, Np\}, where Tc bonds are more polarized than those of Re, Pa and Np.

The order relationships for classes of similar bonds show that there are few classes of poorly polarized bonds, which are less polarized than almost all other classes (hyperedges). They are the heavy alkali metals \{K, Rb, Cs\} and La and Ac. There are also hyperedges of strongly polarized bonds, as those of O and N that are more polarized than almost all the other classes of bonds. Halogens, with the electronegative F, are only more polarized than about half of the other classes of bonds, as the inclusion of not so electronegative elements as I makes that the polarization as a class decreases.

The periodic system of polarized bonds relies on the similarity of chemical elements calculated from their presence in binary compounds. This methodology is chemically general, for it can be extended not only to binary but to $n$-ary compounds. Thus, the method can be applied to any dataset of compounds to assess elemental similarity. The current electronic availability of this information in databases such as Reaxys\textsuperscript{TM} and SciFinder\textsuperscript{TM} make possible the automatisation of the process. Results on the periodic system of chemical elements based on compounds gathered in Reaxys from 1800 up to date are the subject of a forthcoming publication.

The structure of the periodic system here reported frees the concept from the chemical domain and makes it readily applicable to other fields of knowledge. In fact, ordered hypergraphs are also found in information systems and web mining, as recently reported in \cite{He}. Klamt, a decade ago, drew attention to the suitability of hypergraphs for the description of biological, chemical and computational processes \cite{Klamt}. Other examples of systems able to be endowed with similarity subsets are, e.g. ordered systems of countries rated by child development indicators \cite{rainerbook} or by scientific production \cite{Restrepo2014}. Similar examples are found in engineering, hydrology, environmental sciences, sociology, to name but a few areas. Therefore, we envision periodic systems not only of chemical interest but of applicability in other disciplines.

Our results contribute to the undergoing generalisation of network theory to hypergraphs, where the traditional network description as a graph is being abstracted to that of hypergraphs as a mean to model complex relations among multiple entities \cite{He,leal2018formanricci}. We show that hypergraphs can be ordered and that the resulting structure has been at the core of chemistry for more than 150 years.

\vskip6pt

\enlargethispage{20pt}

\section*{Ethics}
No human or animal subjects were involved in this work.

\section*{Data Accessibility}
All calculations on order relationships were performed with the free-ware Python package PyHasse available at \url{https://pyhasse.org/}

\section*{Authors\text{'} Contributions}
WL and GR conceived the study, developed the mathematical formalisation and drafted the manuscript; GR collected and processed the data for the tailored periodic system and wrote the document. Both authors gave final approval for publication

\section*{Competing Interests}
We have no competing interests.

\section*{Acknowledgements}
The authors thank Douglas Klein for motivating this research and Eugen Schwarz and Rainer Br\"uggemann for their valuable comments.

%\disclaimer{Insert disclaimer text here.}

\section*{Appendix}
\begin{definition}
\label{order}
A binary relation $\preceq$ on $X$ is a \emph{partial order} (or order relation) if for all $x,y,z \in X$:
\begin{itemize}
    \item $x \Rightarrow x\preceq x$ (reflexivity)
    \item $x\preceq y$ and $y \preceq x \Rightarrow x=y$ (antisymmetry)
    \item $x \preceq y$ and $y \preceq z \Rightarrow x \preceq z$ (transitivity)
\end{itemize}
The couple $(X,\preceq)$ is called a \emph{partially ordered set}.
\end{definition}

\begin{definition}
\label{similarity}
A binary relation $\sim$ on $X$ is a \emph{similarity relation} (or tolerance relation) if for all $x,y \in X$:
\begin{itemize}
    \item $x \sim x$ (reflexivity)
    \item $x \sim y \Rightarrow y \sim x$ (symmetry)
\end{itemize}
\end{definition}

\begin{definition}
\label{partition}
A family of sets $P$ is a \emph{partition} of $X$ if and only if:
\begin{itemize}
    \item $\varnothing \notin P$
    \item $\cup_{A\in P}A=X$
    \item For all $A,B \in P$, if $A \neq B \Rightarrow A \cap B = \varnothing$
\end{itemize}
\end{definition}

\begin{definition}
\label{HDT-order}
Let $X$ be a non-empty set and $P$ a set of properties $p_i$ of $x\in X$.  For any $x,y \in X$, we say that $x\preceq y$ if $p_i(x) \preceq p_i(y)$, for all $p_i\in P$.
\end{definition}

\begin{definition}
\label{cover-preserving}
Given a partially ordered set $(X,\preceq)$, we say that $x$ \emph{covers} $y$, denoted by $x\preceq : y$, if $x\preceq y$ and there is no $z$ such that $x\preceq z\preceq y$.
A \emph{cover-preserving map} from a partially ordered set $X$ to another $X'$ is a function $f$ such that, if $x,y\in X$ and  $x\preceq : y$, then $f(x)\preceq : f(y)$.
\end{definition}

\begin{definition}
\label{hasse-diagram}
Given a partially ordered set $(X,\preceq)$, its corresponding \emph{Hasse diagram} is a directed graph $(X,E)$, such that $(x,y)\in E$ if $x\preceq : y$, i.e. arrows associated to $(x,y)$ in the directed graph are the cover relations of the partially ordered set $(X,\preceq)$.
%Arrow can also be represented by lines by setting up a convention, e.g. that if $x\preceq y$, then $x$ is placed above $y$ \cite{Trotter}.
\end{definition}

\begin{definition}
\label{hypergraph}
Given a set $X$ and a collection $\{X_i\}_{i \in I}$ of subsets of $X$, a \emph{hypergraph} on $X$ is the pair $H=(X,\{X_i\}_{i \in I})$.
\end{definition}

\begin{definition}
\label{chain}
Given a partially ordered set $(X,\preceq)$, $(X',\preceq)$ is called a \emph{chain} if $X'\subseteq X$ and if for every distinct pair of elements $x,y\in X'$, $x\preceq y$ or $y \preceq x$ holds.
\end{definition}

\begin{definition}
\label{substructure}
Let $H=(X,Y)$ and $H'=(X',Y')$ be hypergraphs.  $H'$ is said to be a \emph{sub-hypergraph} of $H$ if $X' \subseteq X$ and $Y' \subseteq \{X_i\cap X' : X_i \in Y\}$, which is written as $H'\subseteq H$.
\end{definition}

\begin{definition}
\label{hypergraph-isomorphism}
Let $H=(X,\{X_i\}_{i \in I})$ and $H'=(X',\{X'_i\}_{i \in I})$ be hypergraphs on $X$ and $X'$, respectively, they are isomorphic if there exists a bijection $\psi:X \rightarrow X'$ and a permutation $\pi$ of $I$ such that $\psi(X_i)=X_{\pi(i)}'$ .  $\psi$ is called an \emph{isomorphism} and $H$, $H'$ are called isomorphic, denoted by $H\simeq H'$.
\end{definition}

If the elements of $X$ and $X'$ are labelled, the notions of equivalence and equality between hypergraphs arise:

\begin{definition}
\label{hypergraph-equivalence}
Let $H=(X,Y)$ and $H'=(X',Y')$ be isomorphic hypergraphs under $\psi$, they are \emph{equivalent}, denoted by $H\equiv H'$, if $\psi(x)=x'$ and $\psi(X_i)=X_{\pi(i)}'$, where $X_i \in Y$ and $X'_i \in Y'$.  Moreover, if $\pi$ is the identity map, the two hypergraphs are \emph{equal}, denoted by $H = H'$.
\end{definition}

According to Definition \ref{defps}, a periodic system is the couple $(H,\preceq)$ where $H$ is a hypergraph on a set $X$ and $\preceq$ is an order relation.

\begin{definition}
\label{periodic-system-relations}
Let $H=(X,Y)$ and $H'=(X',Y')$ be two hypergraphs such that either $H' \subseteq H$, or $H' \simeq H$ or $H' \equiv H$ or $H' = H$ under $\psi$. If $\psi$ is a cover-preserving map (Definition \ref{cover-preserving}) between $X$ and $X'$, then the periodic systems $PS=(H,\preceq)$ and $PS'=(H',\preceq)$ are $PS'\subseteq PS$, or $PS'\simeq PS$ or $PS'\equiv PS$ or $PS' = PS$, respectively. 
\end{definition}

%%%%%%%%%% Insert bibliography here %%%%%%%%%%%%%%

\clearpage
\section*{Supplementary information}
\textbf{Table S1:} list of dominated, dominating and incomparable bonds for each bond.

\textbf{Dominated bonds.}  The convention is: Bond of element $x$ : (number of dominated bonds + 1, i.e. down set of $x$) : Dominated bonds.

H :( 78 ): Pr, Ni, Yb, Pd, Ru, Na, Nb, Nd, Li, Dy, Y, Tl, Tm, Ra, Rb, Ti, Np, Te, Tb, Tc, Pm, Be, Pa, Ba, Bi, La, As, Po, Pu, Fe, Sr, Ta, Hf, Mo, Ho, Mg, B, H, K, Mn, Lu, Ac, P, Si, Th, U, Sn, Sm, V, Sc, Sb, Es, Os, Eu, Hg, Zn, Co, Cm, Ag, Re, Ca, Ir, Am, Al, Ce, Cd, Ge, Gd, Zr, Ga, In, Cs, Cr, Pb, At, Bk, Cu, Er

Li :( 8 ): Ba, Na, K, Li, Ra, Rb, Sr, Cs

Be :( 43 ): Pr, Th, La, Na, Nd, Li, Sc, Tm, Ra, Rb, Ti, Np, Tb, Pm, Be, Pa, Ba, Yb, Bk, Pu, Ho, Dy, Ta, Hf, Mg, Sr, K, Mn, Lu, Cs, U, Sm, Y, Ac, Cm, Ca, Am, Ce, Gd, Es, Eu, Zr, Er

B :( 67 ): Pr, Ni, Th, La, Na, Nb, Nd, Li, Pb, Y, Tl, Tm, Ra, Rb, Ti, Np, Bk, Tb, Tc, Pm, Be, Pa, Ba, Yb, Bi, Po, Pu, Fe, Dy, Ta, Hf, Hg, Ho, Mg, B, Sr, K, Mn, Lu, Ac, Eu, Si, U, Sn, Sm, V, Sc, Es, Zn, Co, Cm, Ag, Re, Ca, Am, Al, Ce, Cd, Ge, Gd, Zr, Ga, In, Cs, Cr, Cu, Er

C :( 83 ): Pr, Ni, Yb, Pd, Pt, Ru, Na, Nb, Am, Nd, Mg, Li, Pb, Y, Tl, Tm, Ra, Rb, Ti, Np, Te, Rh, Tc, Pm, Be, Pa, Ba, Tb, La, As, Po, W, Pu, Fe, Dy, Ta, Hf, Mo, Ho, C, B, Sr, K, Mn, Lu, Ac, Es, P, Si, Th, U, Sn, Sm, V, Sc, Sb, Bi, Os, Se, Hg, Zn, Co, Cm, Ag, Re, Ca, Ir, Eu, Al, Ce, Cd, Ge, Gd, Au, Zr, Ga, In, Cs, Cr, At, Bk, Cu, Er

N :( 89 ): Ru, Re, Ra, Rb, Rh, Be, Ba, Bi, Bk, Br, P, Os, Es, Hg, Ge, Gd, Ga, Pr, Pt, Pu, C, Pb, Pa, Pd, Cd, Po, Pm, Ho, Hf, K, Mg, Mo, Mn, S, W, Zn, Eu, Zr, Er, Ni, Na, Nb, Nd, Np, Fe, B, Sr, N, Kr, Si, Sn, Sm, V, Sc, Sb, Se, Co, Cm, Ca, Ce, Xe, Lu, Cs, Cr, Cu, La, Li, Tl, Tm, Th, Ti, Te, Tb, Tc, Ta, Yb, Dy, I, U, Y, Ac, Ag, Ir, Am, Al, As, Au, At, In

O :( 91 ): Ru, Re, Ra, Rb, Rh, Be, Ba, Bi, Bk, Br, P, Os, Es, Hg, Ge, Gd, Ga, Pr, Pt, Pu, C, Pb, Pa, Pd, Cd, Po, Pm, Ho, Hf, K, Mg, Mo, Mn, O, S, W, Zn, Eu, Zr, Er, Ni, Na, Nb, Nd, Np, Fe, B, Sr, N, Kr, Si, Sn, Sm, V, Sc, Sb, Se, Co, Cm, Cl, Ca, Ce, Xe, Lu, Cs, Cr, Cu, La, Li, Tl, Tm, Th, Ti, Te, Tb, Tc, Ta, Yb, Dy, I, U, Y, Ac, Ag, Ir, Am, Al, As, Au, At, In

F :( 91 ): Ru, Re, Ra, Rb, Rh, Be, Ba, Bi, Bk, Br, P, Os, Es, Hg, Ge, Gd, Ga, Pr, Pt, Pu, C, Pb, Pa, Pd, Cd, Po, Pm, Ho, Hf, K, Mg, Mo, Mn, S, W, Zn, Eu, Zr, Er, Ni, Na, Nb, Nd, Np, Fe, B, F, Sr, N, Kr, Si, Sn, Sm, V, Sc, Sb, Se, Co, Cm, Cl, Ca, Ce, Xe, Lu, Cs, Cr, Cu, La, Li, Tl, Tm, Th, Ti, Te, Tb, Tc, Ta, Yb, Dy, I, U, Y, Ac, Ag, Ir, Am, Al, As, Au, At, In

Na :( 6 ): Ba, Na, K, Ra, Rb, Cs

Mg :( 33 ): Pr, Th, La, Na, Nd, Tm, Ra, Rb, Tb, Pm, Ba, Yb, Bk, Pu, Ho, Dy, Hf, Mg, Sr, K, Lu, Cs, Sm, Y, Ac, Cm, Ca, Am, Ce, Gd, Eu, Es, Er

Al :( 43 ): Pr, Th, La, Na, Nb, Nd, Li, Sc, Tm, Ra, Rb, Ti, Np, Tb, Pm, Pa, Ba, Yb, Bk, Pu, Ho, Dy, Ta, Hf, Mg, Sr, K, Lu, Eu, U, Sm, Y, Ac, Cm, Ca, Am, Al, Ce, Gd, Es, Cs, Zr, Er

Si :( 56 ): Pr, Th, La, Na, Nb, Nd, Li, Pb, Sc, Tl, Tm, Ra, Rb, Ti, Np, Tb, Tc, Pm, Pa, Ba, Yb, Bk, Pu, Fe, Dy, Ta, Hf, Ho, Mg, Sr, K, Mn, Lu, Ac, Eu, Si, U, Sm, V, Y, Zn, Cm, Re, Ca, Am, Al, Ce, Cd, Gd, Es, Ga, In, Cs, Cr, Zr, Er

P :( 69 ): Pr, Th, La, Na, Nb, Nd, Li, Pb, Y, Tl, Tm, Ra, Rb, Ti, Np, Te, Tb, Tc, Pm, Bi, Pa, Ba, Yb, Ge, Po, Pu, Fe, Dy, Ta, Hf, Mo, Ho, Mg, Sr, K, Mn, Lu, Ac, P, Si, U, Sn, Sm, V, Sc, Sb, Es, Eu, Hg, Zn, Co, Cm, Ag, Re, Ca, Am, Al, Ce, Cd, As, Gd, Zr, Ga, In, Cs, Cr, Bk, Cu, Er

S :( 81 ): Pr, Ni, Yb, Pd, Pt, Ru, Na, Nb, Am, Nd, Li, Pb, Y, Tl, Tm, Ra, Rb, Ti, Np, Te, Rh, Tc, Pm, Bi, Pa, Ba, Tb, La, Si, As, Po, W, Pu, Fe, Dy, Ta, Hf, Mo, Ho, Mg, Sr, K, Mn, Lu, Ac, P, S, Th, U, Sn, Sm, V, Sc, Sb, Es, Os, Se, Hg, Zn, Co, Cm, Ag, Re, Ca, Ir, Eu, Al, Ce, Cd, Ge, Gd, Au, Zr, Ga, In, Cs, Cr, At, Bk, Cu, Er

Cl :( 87 ): Ru, Re, Ra, Rb, Rh, Be, Ba, Bi, Bk, Br, P, Os, Es, Hg, Ge, Gd, Ga, Pr, Pt, Pu, Pb, Pa, Pd, Cd, Po, Pm, Ho, Hf, K, Mg, Mo, Mn, S, W, Zn, Eu, Zr, Er, Ni, Na, Nb, Nd, Np, Fe, Sr, Kr, Si, Sn, Sm, V, Sc, Sb, Se, Co, Cm, Cl, Ca, Ce, Xe, Lu, Cs, Cr, Cu, La, Li, Tl, Tm, Th, Ti, Te, Tb, Tc, Ta, Yb, Dy, I, U, Y, Ac, Ag, Ir, Am, Al, As, Au, At, In

K :( 3 ): Cs, K, Rb

Ca :( 7 ): Ba, Sr, K, Ra, Rb, Cs, Ca

Sc :( 35 ): Pr, Th, La, Na, Nd, Sc, Tm, Ra, Rb, Np, Tb, Pm, Ba, Yb, Bk, Pu, Ho, Dy, Hf, Sr, K, Lu, Eu, Sm, Y, Ac, Cm, Ca, Am, Ce, Gd, Es, Cs, Zr, Er

Ti :( 40 ): Pr, Th, La, Na, Nd, Sc, Tm, Ra, Rb, Ti, Np, Tb, Pm, Pa, Ba, Yb, Bk, Pu, Ho, Dy, Ta, Hf, Mg, Sr, K, Lu, Eu, U, Sm, Y, Ac, Cm, Ca, Am, Ce, Gd, Es, Cs, Zr, Er

V :( 43 ): Pr, Th, La, Na, Nb, Nd, Sc, Tl, Tm, Ra, Rb, Ti, Np, Tb, Pm, Pa, Ba, Yb, Bk, Pu, Ho, Dy, Ta, Hf, Mg, Sr, K, Lu, Eu, U, Sm, V, Y, Ac, Cm, Ca, Am, Ce, Gd, Es, Cs, Zr, Er
Cr :( 46 ): Pr, Th, La, Na, Nb, Nd, Li, Sc, Tl, Tm, Ra, Rb, Ti, Np, Tb, Pm, Pa, Ba, Yb, Bk, Pu, Ho, Dy, Ta, Hf, Mg, Sr, K, Lu, Eu, U, Sm, V, Y, Ac, Cm, Ca, Am, Al, Ce, Gd, Es, Cs, Cr, Zr, Er

Mn :( 42 ): Pr, Th, La, Na, Nd, Li, Sc, Tm, Ra, Rb, Ti, Np, Tb, Pm, Pa, Ba, Yb, Bk, Pu, Ho, Dy, Ta, Hf, Mg, Sr, K, Mn, Lu, Eu, U, Sm, Y, Ac, Cm, Ca, Am, Ce, Gd, Es, Cs, Zr, Er

Fe :( 52 ): Pr, Th, La, Na, Nb, Nd, Li, Sc, Tl, Tm, Ra, Rb, Ti, Np, Tb, Pm, Pa, Ba, Yb, Bk, Pu, Fe, Dy, Ta, Hf, Ho, Mg, Sr, K, Mn, Lu, Ac, Eu, U, Sm, V, Y, Zn, Cm, Ca, Am, Al, Ce, Cd, Gd, Es, Ga, In, Cs, Cr, Zr, Er

Co :( 54 ): Pr, Th, La, Na, Nb, Nd, Li, Pb, Sc, Tl, Tm, Ra, Rb, Ti, Np, Tb, Pm, Pa, Ba, Yb, Bk, Pu, Fe, Dy, Ta, Hf, Ho, Mg, Sr, K, Mn, Lu, Ac, Eu, U, Sm, V, Y, Zn, Co, Cm, Ca, Am, Al, Ce, Cd, Gd, Es, Ga, In, Cs, Cr, Zr, Er

Ni :( 59 ): Pr, Ni, Th, La, Na, Nb, Nd, Li, Pb, Sc, Tl, Tm, Ra, Rb, Ti, Np, Tb, Tc, Pm, Pa, Ba, Yb, Bk, Pu, Fe, Dy, Ta, Hf, Ho, Mg, Sr, K, Mn, Lu, Ac, Eu, Si, U, Sm, V, Y, Es, Zn, Co, Cm, Re, Ca, Am, Al, Ce, Cd, Gd, Cu, Ga, In, Cs, Cr, Zr, Er

Cu :( 57 ): Pr, Th, La, Na, Nb, Nd, Li, Pb, Sc, Tl, Tm, Ra, Rb, Ti, Np, Tb, Tc, Pm, Pa, Ba, Yb, Bk, Pu, Fe, Dy, Ta, Hf, Ho, Mg, Sr, K, Mn, Lu, Ac, Eu, Si, U, Sm, V, Y, Es, Zn, Cm, Re, Ca, Am, Al, Ce, Cd, Gd, Cu, Ga, In, Cs, Cr, Zr, Er

Zn :( 47 ): Pr, Th, La, Na, Nb, Nd, Li, Sc, Tl, Tm, Ra, Rb, Ti, Np, Tb, Pm, Pa, Ba, Yb, Bk, Pu, Ho, Dy, Ta, Hf, Mg, Sr, K, Mn, Lu, Ac, Eu, U, Sm, V, Y, Zn, Cm, Ca, Am, Al, Ce, Gd, Es, Cs, Zr, Er

Ga :( 48 ): Pr, Th, La, Na, Nb, Nd, Li, Sc, Tl, Tm, Ra, Rb, Ti, Np, Tb, Pm, Pa, Ba, Yb, Bk, Pu, Ho, Dy, Ta, Hf, Mg, Sr, K, Lu, Eu, U, Sm, V, Y, Ac, Cm, Ca, Am, Al, Ce, Cd, Gd, Es, Ga, In, Cs, Zr, Er

Ge :( 57 ): Pr, Th, La, Na, Nb, Nd, Li, Pb, Sc, Tl, Tm, Ra, Rb, Ti, Np, Bk, Tb, Tc, Pm, Pa, Ba, Yb, Po, Pu, Ho, Dy, Ta, Hf, Hg, Mg, Sr, K, Lu, Eu, U, Sn, Sm, V, Y, Ac, Cm, Ag, Re, Ca, Am, Al, Ce, Cd, Ge, Gd, Es, Ga, In, Cs, Cr, Zr, Er

As :( 62 ): Pr, Th, La, Na, Nb, Nd, Li, Pb, Sc, Tl, Tm, Ra, Rb, Ti, Np, Te, Tb, Tc, Pm, Bi, Pa, Ba, Yb, Ge, Po, Pu, Ho, Dy, Ta, Hf, K, Mg, Sr, Mo, Lu, Bk, Eu, U, Sn, Sm, V, Y, Sb, Hg, Ac, Cm, Ag, Re, Ca, Am, Al, Ce, Cd, As, Gd, Es, Ga, In, Cs, Cr, Zr, Er

Se :( 76 ): Pr, Ru, Pd, Pt, La, Na, Nb, Nd, Li, Pb, Sc, Tl, Tm, Ra, Rb, Ti, Np, Te, Rh, Tc, Pm, Bi, Pa, Ba, Tb, Yb, Ge, Po, W, Pu, Fe, Dy, Ta, Hf, Mo, Ho, Mg, Sr, K, Mn, Lu, Ac, Eu, Si, Th, U, Sn, Sm, V, Y, Sb, Es, Os, Se, Hg, Zn, Cm, Ag, Re, Ca, Ir, Am, Al, Ce, Cd, As, Gd, Au, At, Ga, In, Cs, Cr, Bk, Zr, Er

Br :( 79 ): Pr, Ru, Pd, Pt, La, Na, Nb, Nd, Li, Pb, Sc, Tl, Tm, Ra, Rb, Ti, Np, Te, Rh, Tc, Pm, Bi, Xe, Pa, Ba, Tb, Yb, Ge, Po, W, Pu, Fe, Br, Dy, Ta, Hf, Mo, Ho, Mg, I, Sr, K, Mn, Lu, Ac, Eu, Si, Th, U, Sn, Sm, V, Y, Sb, Es, Os, Se, Hg, Zn, Cm, Ag, Re, Ca, Ir, Am, Al, Ce, Cd, As, Gd, Au, At, Ga, In, Cs, Cr, Bk, Zr, Er

Kr :( 76 ): Pr, Ru, Pd, Pt, La, Na, Nb, Nd, Li, Pb, Sc, Tl, Tm, Ra, Rb, Ti, Np, Te, Rh, Tc, Pm, Bi, Xe, Pa, Ba, Tb, Yb, Ge, Po, W, Pu, Ho, Dy, Ta, Hf, K, Mg, I, Sr, Mo, Mn, Lu, Ac, Kr, Th, U, Sn, Sm, V, Y, Sb, Es, Os, Eu, Hg, Zn, Cm, Ag, Re, Ca, Ir, Am, Al, Ce, Cd, As, Gd, Au, At, Ga, In, Cs, Cr, Bk, Zr, Er

Rb :( 2 ): Cs, Rb

Sr :( 6 ): Ba, Sr, K, Ra, Rb, Cs

Y :( 21 ): Pr, Ac, Sm, Ba, La, Sr, K, Yb, Nd, Eu, Ce, Am, Cs, Gd, Ra, Rb, Dy, Y, Tb, Ca, Pm

Zr :( 32 ): Pr, Th, La, Na, Nd, Tm, Ra, Rb, Tb, Pm, Ba, Yb, Bk, Ho, Dy, Pu, Sr, K, Lu, Cs, Sm, Y, Ac, Cm, Ca, Am, Ce, Gd, Es, Eu, Zr, Er

Nb :( 38 ): Pr, Th, La, Na, Nb, Nd, Sc, Tm, Ra, Rb, Np, Tb, Pm, Pa, Ba, Yb, Bk, Pu, Ho, Dy, Hf, Sr, K, Lu, Eu, U, Sm, Y, Ac, Cm, Ca, Am, Ce, Gd, Es, Cs, Zr, Er

Mo :( 48 ): Pr, Th, La, Na, Nb, Nd, Pb, Sc, Tl, Tm, Ra, Rb, Np, Tb, Po, Pm, Pa, Ba, Yb, Bi, Bk, Pu, Ho, Dy, Ta, Hf, K, Mg, Sr, Mo, Lu, Eu, U, Sn, Sm, Y, Sb, Ac, Cm, Ca, Am, Ce, Gd, Es, In, Cs, Zr, Er

Tc :( 49 ): Pr, Th, La, Na, Nb, Nd, Li, Pb, Sc, Tl, Tm, Ra, Rb, Ti, Np, Tb, Tc, Pm, Pa, Ba, Yb, Bk, Pu, Ho, Dy, Ta, Hf, Mg, Sr, K, Lu, Eu, U, Sm, V, Y, Ac, Cm, Re, Ca, Am, Ce, Cd, Gd, Es, In, Cs, Zr, Er

Ru :( 61 ): Pr, Ru, Th, La, Na, Nb, Nd, Li, Pb, Sc, Tl, Tm, Ra, Rb, Ti, Np, Te, Tb, Tc, Pm, Pa, Ba, Yb, Bi, Po, Pu, Ho, Dy, Ta, Hf, K, Mg, Sr, Mo, Lu, Eu, U, Sn, Sm, V, Y, Sb, Es, Os, Hg, Ac, Cm, Ag, Re, Ca, Am, Al, Ce, Cd, Gd, At, In, Cs, Bk, Zr, Er

Rh :( 62 ): Pr, Ru, Th, La, Na, Nb, Nd, Li, Pb, Sc, Tl, Tm, Ra, Rb, Ti, Np, Te, Rh, Tc, Pm, Pa, Ba, Tb, Yb, Bi, Po, Pu, Ho, Dy, Ta, Hf, K, Mg, Sr, Mo, Lu, Eu, U, Sn, Sm, V, Y, Sb, Es, Os, Hg, Ac, Cm, Ag, Re, Ca, Am, Al, Ce, Cd, Gd, At, In, Cs, Bk, Zr, Er

Pd :( 67 ): Pr, Ru, Pd, La, Na, Nb, Nd, Li, Pb, Sc, Tl, Tm, Ra, Rb, Ti, Np, Te, Tb, Tc, Pm, Bi, Pa, Ba, Yb, Ge, Po, Pu, Ho, Dy, Ta, Hf, K, Mg, Sr, Mo, Lu, Bk, Eu, Th, U, Sn, Sm, V, Y, Sb, Es, Os, Hg, Ac, Cm, Ag, Re, Ca, Ir, Am, Al, Ce, Cd, As, Gd, At, Ga, In, Cs, Cr, Zr, Er

Ag :( 50 ): Pr, Th, La, Na, Nb, Nd, Li, Pb, Sc, Tl, Tm, Ra, Rb, Ti, Np, Tb, Tc, Pm, Pa, Ba, Yb, Bk, Pu, Ho, Dy, Ta, Hf, Mg, Sr, K, Lu, Eu, U, Sm, V, Y, Ac, Cm, Ag, Re, Ca, Am, Ce, Cd, Gd, Es, In, Cs, Zr, Er

Cd :( 43 ): Pr, Th, La, Na, Nb, Nd, Sc, Tl, Tm, Ra, Rb, Ti, Np, Tb, Pm, Pa, Ba, Yb, Bk, Pu, Ho, Dy, Ta, Hf, Mg, Sr, K, Lu, Eu, U, Sm, Y, Ac, Cm, Ca, Am, Ce, Cd, Gd, Es, Cs, Zr, Er

In :( 41 ): Pr, Th, La, Na, Nb, Nd, Sc, Tl, Tm, Ra, Rb, Np, Tb, Pm, Pa, Ba, Yb, Bk, Pu, Ho, Dy, Ta, Hf, Sr, K, Lu, Eu, U, Sm, Y, Ac, Cm, Ca, Am, Ce, Gd, Es, In, Cs, Zr, Er

Sn :( 43 ): Pr, Th, La, Na, Nb, Nd, Pb, Sc, Tl, Tm, Ra, Rb, Np, Tb, Pm, Pa, Ba, Yb, Bk, Pu, Ho, Dy, Ta, Hf, Sr, K, Lu, Eu, U, Sn, Sm, Y, Ac, Cm, Ca, Am, Ce, Gd, Es, In, Cs, Zr, Er

Sb :( 46 ): Pr, Th, La, Na, Nb, Nd, Pb, Sc, Tl, Tm, Ra, Rb, Np, Tb, Po, Pm, Pa, Ba, Yb, Bi, Bk, Pu, Ho, Dy, Ta, Hf, Sr, K, Lu, Eu, U, Sn, Sm, Y, Sb, Ac, Cm, Ca, Am, Ce, Gd, Es, In, Cs, Zr, Er

Te :( 50 ): Pr, Th, La, Na, Nb, Nd, Pb, Sc, Tl, Tm, Ra, Rb, Ti, Np, Te, Tb, Po, Pm, Pa, Ba, Yb, Bi, Bk, Pu, Ho, Dy, Ta, Hf, Mg, Sr, K, Lu, Eu, U, Sn, Sm, Y, Sb, Ac, Cm, Ca, Am, Ce, Cd, Gd, Es, In, Cs, Zr, Er

I :( 57 ): Pr, W, La, Na, Nb, Nd, Li, Pb, Sc, Tl, Tm, Ra, Rb, Ti, Np, Te, Tb, Po, Pm, Pa, Ba, Yb, Bi, Bk, Pu, Ho, Dy, Ta, Hf, K, Mg, I, Sr, Mo, Lu, Eu, Th, U, Sn, Sm, V, Y, Sb, Es, Hg, Ac, Cm, Ca, Am, Ce, Cd, Gd, At, In, Cs, Zr, Er

Xe :( 58 ): Pr, W, La, Na, Nb, Nd, Li, Pb, Sc, Tl, Tm, Ra, Rb, Ti, Np, Te, Tb, Po, Pm, Pa, Ba, Yb, Bi, Bk, Cd, Pu, Ho, Dy, Ta, Hf, K, Mg, Sr, Mo, Lu, Eu, Th, U, Sn, Sm, V, Y, Sb, Es, Hg, Ac, Cm, Re, Ca, Am, Ce, Xe, Gd, At, In, Cs, Zr, Er

Cs :( 1 ): Cs

Ba :( 4 ): Cs, K, Ba, Rb

La :( 8 ): Ac, Ba, La, Sr, K, Ra, Rb, Cs

Ce :( 12 ): Ac, Ba, La, Sr, K, Yb, Ce, Ra, Rb, Cs, Tb, Ca

Pr :( 9 ): Pr, Ac, Ba, La, Sr, K, Ra, Rb, Cs
Nd :( 10 ): Pr, Ac, Ba, La, Sr, K, Nd, Ra, Rb, Cs

Pm :( 10 ): Pr, Ac, Ba, La, Sr, K, Ra, Rb, Cs, Pm

Sm :( 12 ): Pr, Ac, Sm, Ba, La, Sr, K, Nd, Ra, Rb, Cs, Pm

Eu :( 17 ): Pr, Ac, Sm, Ba, La, Sr, K, Yb, Nd, Eu, Gd, Ra, Rb, Cs, Tb, Ca, Pm

Gd :( 15 ): Pr, Ac, Sm, Ba, La, Sr, K, Yb, Nd, Gd, Ra, Rb, Cs, Ca, Pm

Tb :( 11 ): Ac, Ba, La, Sr, K, Yb, Ra, Rb, Cs, Tb, Ca

Dy :( 18 ): Pr, Ac, Sm, Ba, La, Sr, K, Yb, Nd, Eu, Gd, Ra, Rb, Dy, Cs, Tb, Ca, Pm
Ho :( 20 ): Pr, Ra, Sm, Ba, La, Sr, K, Yb, Nd, Ac, Am, Eu, Gd, Ho, Rb, Dy, Cs, Tb, Ca, Pm

Er :( 21 ): Pr, Ra, Sm, Ba, La, Sr, K, Yb, Nd, Ac, Am, Eu, Gd, Ho, Rb, Dy, Cs, Er, Tb, Ca, Pm

Tm :( 22 ): Pr, La, Nd, Tm, Ra, Rb, Tb, Pm, Ba, Yb, Ho, Dy, Sr, K, Cs, Sm, Ac, Ca, Am, Gd, Eu, Er

Yb :( 10 ): Ac, Ba, La, Sr, K, Yb, Ra, Rb, Cs, Ca

Lu :( 25 ): Pr, La, Nd, Lu, Ra, Rb, Tb, Pm, Ba, Yb, Ho, Dy, Sr, K, Cs, Sm, Y, Ac, Ca, Am, Ce, Gd, Eu, Tm, Er

Hf :( 32 ): Pr, Th, La, Na, Nd, Tm, Ra, Rb, Tb, Pm, Ba, Yb, Bk, Pu, Ho, Dy, Hf, Sr, K, Lu, Cs, Sm, Y, Ac, Cm, Ca, Am, Ce, Gd, Eu, Es, Er

Ta :( 38 ): Pr, Th, La, Na, Nd, Sc, Tm, Ra, Rb, Np, Tb, Pm, Pa, Ba, Yb, Bk, Pu, Ho, Dy, Ta, Hf, Sr, K, Lu, Eu, U, Sm, Y, Ac, Cm, Ca, Am, Ce, Gd, Es, Cs, Zr, Er

W :( 50 ): Pr, W, La, Na, Nb, Nd, Pb, Sc, Tl, Tm, Ra, Rb, Np, Tb, Po, Pm, Pa, Ba, Yb, Bi, Bk, Pu, Ho, Dy, Ta, Hf, K, Mg, Sr, Mo, Lu, Eu, U, Sn, Sm, Y, Sb, Es, Th, Ac, Cm, Ca, Am, Ce, Gd, At, In, Cs, Zr, Er

Re :( 48 ): Pr, Th, La, Na, Nb, Nd, Li, Pb, Sc, Tl, Tm, Ra, Rb, Ti, Np, Tb, Pm, Pa, Ba, Yb, Bk, Pu, Ho, Dy, Ta, Hf, Mg, Sr, K, Lu, Eu, U, Sm, V, Y, Ac, Cm, Re, Ca, Am, Ce, Cd, Gd, Es, In, Cs, Zr, Er

Os :( 57 ): Pr, Th, La, Na, Nb, Nd, Li, Pb, Sc, Tl, Tm, Ra, Rb, Ti, Np, Te, Tb, Po, Pm, Pa, Ba, Yb, Bi, Bk, Pu, Ho, Dy, Ta, Hf, K, Mg, Sr, Mo, Lu, Eu, U, Sn, Sm, V, Y, Sb, Es, Os, Hg, Ac, Cm, Re, Ca, Am, Ce, Cd, Gd, At
In, Cs, Zr, Er

Ir :( 64 ): Pr, Ru, Th, La, Na, Nb, Nd, Li, Pb, Sc, Tl, Tm, Ra, Rb, Ti, Np, Te, Tb, Tc, Pm, Pa, Ba, Yb, Bi, Po, Pu, Ho, Dy, Ta, Hf, K, Mg, Sr, Mo, Lu, Bk, Eu, U, Sn, Sm, V, Y, Sb, Es, Os, Hg, Ac, Cm, Ag, Re, Ca, Ir, Am, Al, Ce, Cd, Gd, At, Ga, In, Cs, Cr, Zr, Er

Pt :( 64 ): Pr, Ru, Th, Pt, La, Na, Nb, Nd, Li, Pb, Sc, Tl, Tm, Ra, Rb, Ti, Np, Te, Rh, Tc, Pm, Pa, Ba, Tb, Yb, Bi, Po, Pu, Ho, Dy, Ta, Hf, K, Mg, Sr, Mo, Lu, Eu, U, Sn, Sm, V, Y, Sb, Es, Os, Hg, Ac, Cm, Ag, Re, Ca, Am, Al, Ce, Cd, Gd, At, Ga, In, Cs, Bk, Zr, Er

Au :( 65 ): Pr, Ru, W, La, Na, Nb, Nd, Li, Pb, Sc, Tl, Tm, Ra, Rb, Ti, Np, Te, Rh, Tc, Pm, Pa, Ba, Tb, Yb, Bi, Po, Pu, Ho, Dy, Ta, Hf, K, Mg, Sr, Mo, Lu, Eu, Th, U, Sn, Sm, V, Y, Sb, Es, Os, Hg, Ac, Cm, Ag, Re, Ca, Am, Al, Ce, Cd, Gd, Au, At, Ga, In, Cs, Bk, Zr, Er

Hg :( 50 ): Pr, Th, La, Na, Nb, Nd, Li, Pb, Sc, Tl, Tm, Ra, Rb, Ti, Np, Tb, Po, Pm, Pa, Ba, Yb, Bk, Pu, Ho, Dy, Ta, Hf, Hg, Mg, Sr, K, Lu, Eu, U, Sn, Sm, V, Y, Ac, Cm, Ca, Am, Ce, Cd, Gd, Es, In, Cs, Zr, Er

Tl :( 40 ): Pr, Th, La, Na, Nb, Nd, Sc, Tl, Tm, Ra, Rb, Np, Tb, Pm, Pa, Ba, Yb, Bk, Pu, Ho, Dy, Ta, Hf, Sr, K, Lu, Eu, U, Sm, Y, Ac, Cm, Ca, Am, Ce, Gd, Es, Cs, Zr, Er

Pb :( 41 ): Pr, Th, La, Na, Nb, Nd, Pb, Sc, Tl, Tm, Ra, Rb, Np, Tb, Pm, Pa, Ba, Yb, Bk, Pu, Ho, Dy, Ta, Hf, Sr, K, Lu, Eu, U, Sm, Y, Ac, Cm, Ca, Am, Ce, Gd, Es, Cs, Zr, Er

Bi :( 37 ): Pr, Th, La, Na, Nd, Tm, Ra, Rb, Np, Tb, Pm, Pa, Ba, Yb, Bi, Bk, Pu, Ho, Dy, Hf, Sr, K, Lu, Cs, U, Sm, Y, Ac, Cm, Ca, Am, Ce, Gd, Es, Eu, Zr, Er

Po :( 40 ): Pr, Th, La, Na, Nb, Nd, Sc, Tm, Ra, Rb, Np, Tb, Po, Pm, Pa, Ba, Yb, Bk, Pu, Ho, Dy, Ta, Hf, Sr, K, Lu, Eu, U, Sm, Y, Ac, Cm, Ca, Am, Ce, Gd, Es, Cs, Zr, Er

At :( 40 ): Pr, Th, La, Na, Nb, Nd, Sc, Tm, Ra, Rb, Np, Tb, Pm, Pa, Ba, Yb, Bi, Bk, Pu, Ho, Dy, Hf, Sr, K, Lu, Eu, U, Sm, Y, Es, Ac, Cm, Ca, Am, Ce, Gd, At, Cs, Zr, Er

Ra :( 3 ): Cs, Ra, Rb

Ac :( 6 ): Ac, Ba, K, Ra, Rb, Cs

Th :( 10 ): Pr, Ac, Ba, La, Sr, K, Ra, Rb, Cs, Th

Pa :( 20 ): Pr, Ac, Pu, Sm, Ba, La, Sr, K, Yb, Nd, Np, Pa, Gd, U, Ra, Rb, Cs, Th, Ca, Pm

U :( 18 ): Pr, Ac, Pu, Sm, Ba, La, Sr, K, Yb, Nd, Np, U, Ra, Rb, Cs, Th, Ca, Pm

Np :( 16 ): Pr, Ac, Sm, Ba, La, Sr, K, Pu, Nd, Np, Ra, Rb, Cs, Th, Ca, Pm

Pu :( 13 ): Pr, Ac, Sm, Ba, La, Sr, K, Pu, Nd, Ra, Rb, Cs, Pm

Am :( 14 ): Pr, Ac, Ba, La, Sr, K, Yb, Am, Ra, Rb, Cs, Tb, Ca, Pm

Cm :( 22 ): Pr, La, Nd, Ra, Rb, Tb, Pm, Ba, Yb, Ho, Dy, Pu, Sr, K, Cs, Sm, Ac, Cm, Ca, Am, Gd, Eu

Bk :( 20 ): Pr, Th, Ac, Sm, Ba, La, Sr, K, Yb, Nd, Bk, Eu, Gd, Pu, Ra, Rb, Cs, Tb, Ca, Pm

Es :( 26 ): Pr, Rb, La, Nd, Ra, Th, Tb, Pm, Ba, Yb, Bk, Ho, Dy, Pu, Sr, K, Cs, Sm, Ac, Cm, Ca, Am, Gd, Eu, Es, Er

\textbf{Dominating bonds.}  The convention is: Bond of element $x$ : (number of dominating bonds + 1, i.e. up set of $x$) : Dominating bonds.

\textbf{upsets} 

H :( 1 ): H

Li :( 39 ): Ni, Pt, Ru, Li, Re, Pd, Rh, Tc, Be, Si, As, Fe, Br, Hg, C, B, F, I, H, Mn, O, N, P, S, Kr, Os, Se, Zn, Co, Ag, Cl, Ir, Al, Xe, Ge, Au, Ga, Cr, Cu

Be :( 8 ): Be, C, B, F, H, Cl, O, N

B :( 6 ): C, B, F, H, O, N

C :( 4 ): C, F, O, N

N :( 3 ): F, O, N

O :( 1 ): O

F :( 1 ): F

Na :( 60 ): Ni, Pt, Ru, Na, Nb, Mg, Li, Pb, Re, Tl, Pd, Ti, Te, Rh, Tc, Ta, Be, Xe, Bi, Si, As, Po, Fe, Br, Hf, Hg, At, C, B, F, I, H, Mo, Mn, O, N, P, S, Kr, W, V, Sc, Sb, Os, Se, Zn, Co, Ag, Cl, Ir, Al, Cd, Ge, Au, Zr, Ga, In, Cr, Cu, Sn

Mg :( 45 ): Ni, Pt, Ru, Mg, Re, Pd, Ti, Te, Rh, Tc, Be, Xe, Si, As, Fe, Br, Hg, C, B, F, I, H, Mo, Mn, O, N, P, S, Kr, W, V, Os, Se, Zn, Co, Ag, Cl, Ir, Al, Cd, Ge, Au, Ga, Cr, Cu

Al :( 29 ): Ni, Pt, Ru, Pd, Rh, Si, As, Fe, Br, C, B, F, H, O, N, P, S, Kr, Se, Zn, Co, Cl, Ir, Al, Ge, Au, Ga, Cr, Cu

Si :( 14 ): Ni, C, B, F, H, Cl, S, O, N, P, Si, Se, Br, Cu

P :( 8 ): C, F, H, Cl, O, N, P, S

S :( 5 ): Cl, S, F, O, N

Cl :( 3 ): Cl, O, F

K :( 90 ): Ru, Re, Rh, Be, Ba, Bi, Bk, Br, H, P, Os, Es, Hg, Ge, Gd, Ga, Pr, Pt, Pu, C, Pb, Pa, Pd, Cd, Po, Pm, Ho, Hf, K, Mg, Mo, Mn, O, S, W, Zn, Eu, Zr, Er, Ni, Na, Nb, Nd, Np, Fe, B, F, Sr, N, Kr, Si, Sn, Sm, V, Sc, Sb, Se, Co, Cm, Cl, Ca, Ce, Xe, Lu, Cr, Cu, La, Li, Tl, Tm, Th, Ti, Te, Tb, Tc, Ta, Yb, Dy, I, U, Y, Ac, Ag, Ir, Am, Al, As, Au, At, In

Ca :( 77 ): Ni, Pt, Ru, Nb, Mg, Bi, Pb, Y, Tl, Tm, Pd, Ti, Np, Te, Rh, Tc, Ta, Be, Xe, Pa, U, Tb, Yb, Si, As, Po, Fe, Br, Dy, Hf, Hg, Ho, C, B, F, I, H, Mo, Mn, Lu, Bk, O, N, P, S, Re, Kr, W, V, Sc, Sb, Os, Se, Es, Zn, Co, Cm, Ag, Cl, At, Ca, Ir, Am, Al, Ce, Cd, Ge, Gd, Au, Zr, Ga, In, Eu, Cr, Er, Cu, Sn

Sc :( 54 ): Ni, Pt, Ru, Nb, Pb, Re, Tl, Pd, Ti, Te, Rh, Tc, Ta, Be, Xe, Si, As, Po, Fe, Br, Hg, C, B, F, I, H, Mo, Mn, O, N, P, S, Kr, W, V, Sc, Sb, Os, Se, Zn, Co, Ag, Cl, Ir, Al, Cd, Ge, Au, At, Ga, In, Cr, Cu, Sn

Ti :( 42 ): Ni, Pt, Ru, Re, Pd, Ti, Te, Rh, Tc, Be, Xe, Si, As, Fe, Br, Hg, C, B, F, I, H, Mn, O, N, P, S, Kr, V, Os, Se, Zn, Co, Ag, Cl, Ir, Al, Cd, Ge, Au, Ga, Cr, Cu

V :( 36 ): Ni, Pt, Ru, Re, Pd, Rh, Tc, As, Fe, Br, Hg, C, B, F, I, H, O, N, P, S, Kr, V, Os, Se, Zn, Co, Ag, Cl, Ir, Si, Xe, Ge, Au, Ga, Cr, Cu

Cr :( 22 ): Ni, Pd, As, Fe, Br, C, B, P, F, H, O, N, Kr, S, Se, Co, Cl, Ir, Si, Ge, Cr, Cu

Mn :( 20 ): Be, C, B, Co, Br, Zn, F, H, Kr, Cl, S, O, N, P, Si, Se, Fe, Mn, Ni, Cu

Fe :( 16 ): Ni, C, B, Co, F, H, Cl, S, O, N, P, Si, Se, Fe, Br, Cu

Co :( 11 ): Ni, C, B, Co, F, H, Cl, O, N, P, S

Ni :( 9 ): Ni, C, B, F, H, Cl, O, N, S

Cu :( 11 ): Ni, C, B, F, H, Cl, O, N, P, S, Cu

Zn :( 18 ): Ni, C, B, Co, Zn, F, H, Kr, Cl, S, O, N, P, Si, Se, Fe, Br, Cu

Ga :( 24 ): Ni, Pt, Pd, As, Fe, Br, C, B, P, F, H, O, N, Kr, S, Se, Co, Cl, Ir, Si, Ge, Au, Ga, Cu

Ge :( 15 ): As, C, B, Pd, F, H, Cl, Ge, O, N, P, S, Kr, Br, Se

As :( 13 ): C, Pd, F, H, Cl, As, O, N, P, S, Kr, Br, Se

Se :( 8 ): C, Cl, F, O, N, S, Br, Se

Br :( 5 ): Cl, F, Br, O, N

Kr :( 5 ): Cl, F, Kr, O, N

Rb :( 92 ): Ru, Re, Ra, Rb, Rh, Be, Ba, Bi, Bk, Br, H, P, Os, Es, Hg, Ge, Gd, Ga, Pr, Pt, Pu, C, Pb, Pa, Pd, Cd, Po, Pm, Ho, Hf, K, Mg, Mo, Mn, O, S, W, Zn, Eu, Zr, Er, Ni, Na, Nb, Nd, Np, Fe, B, F, Sr, N, Kr, Si, Sn, Sm, V, Sc, Sb, Se, Co, Cm, Cl, Ca, Ce, Xe, Lu, Cr, Cu, La, Li, Tl, Tm, Th, Ti, Te, Tb, Tc, Ta, Yb, Dy, I, U, Y, Ac, Ag, Ir, Am, Al, As, Au, At, In

Sr :( 86 ): Ru, Re, Rh, Be, Bi, Bk, Br, H, P, Os, Ge, Gd, Ga, Pr, Pt, Pu, C, Pb, Pa, Pd, Cd, Po, Pm, Ho, Hf, Hg, Mg, Mo, Mn, O, Zr, S, W, Zn, Eu, Es, Er, Ni, Nb, Nd, Np, Fe, B, F, Sr, N, Kr, Si, Sn, Sm, V, Sc, Sb, Se, Co, Cm, Cl, Ca, Ce, Xe, Lu, Cr, Cu, La, Li, Tl, Tm, Th, Ti, Te, Tb, Tc, Ta, Yb, Dy, I, U, Y, Ag, Ir, Am, Al, As, Au, At, In

Y :( 60 ): Ni, Pt, Ru, Nb, Mg, Bi, Pb, Y, Tl, Lu, Pd, Ti, Te, Rh, Tc, Ta, Be, Xe, Si, As, Po, Fe, Br, Hf, Hg, At, C, B, F, I, H, Mo, Mn, O, N, P, S, Re, Kr, W, V, Sc, Sb, Os, Se, Zn, Co, Ag, Cl, Ir, Al, Cd, Ge, Au, Zr, Ga, In, Cr, Cu, Sn

Zr :( 56 ): Ni, Pt, Ru, Nb, Bi, Pb, Re, Tl, Pd, Ti, Te, Rh, Tc, Ta, Be, Xe, Si, As, Po, Fe, Br, Hg, At, C, B, F, I, H, Mo, Mn, O, N, P, S, Kr, W, V, Sc, Sb, Os, Se, Zn, Co, Ag, Cl, Ir, Al, Cd, Ge, Au, Zr, Ga, In, Cr, Cu, Sn

Nb :( 49 ): Ni, Pt, Ru, Nb, Pb, Re, Tl, Pd, Te, Rh, Tc, Xe, Si, As, Po, Fe, Br, Hg, C, B, F, I, H, Mo, O, N, P, S, Kr, W, V, Sb, Os, Se, Zn, Co, Ag, Cl, Ir, Al, Cd, Ge, Au, At, Ga, In, Cr, Cu, Sn

Mo :( 23 ): Ru, Pt, Pd, Rh, Se, Br, C, F, I, H, Mo, O, N, P, S, W, Os, Kr, Cl, Ir, Xe, As, Au

Tc :( 25 ): Ni, Pt, Ru, Pd, Rh, Tc, As, Br, C, B, P, F, H, O, N, Kr, S, Se, Ag, Cl, Ir, Si, Ge, Au, Cu

Ru :( 16 ): Ru, C, Ir, Pd, Pt, F, H, Cl, O, N, Kr, S, Au, Br, Rh, Se

Rh :( 12 ): C, Pt, Cl, F, O, N, Kr, S, Au, Br, Rh, Se

Pd :( 11 ): C, Pd, F, H, Cl, O, N, Kr, S, Br, Se

Ag :( 21 ): As, C, B, Pd, Ag, F, H, Cl, Ge, O, N, P, S, Ru, Kr, Br, Au, Pt, Rh, Se, Ir

Cd :( 35 ): Ni, Pt, Ru, Re, Pd, Te, Rh, Tc, Xe, As, Fe, Br, Hg, C, B, P, F, I, H, O, N, Kr, S, Os, Se, Co, Ag, Cl, Ir, Si, Cd, Ge, Au, Ga, Cu

In :( 39 ): Ni, Pt, Ru, Re, Pd, Te, Rh, Tc, As, Fe, Br, Hg, C, B, P, F, I, H, Mo, O, N, Kr, S, Sn, W, Sb, Os, Se, Co, Ag, Cl, Ir, Si, Xe, Ge, Au, Ga, In, Cu

Sn :( 29 ): Ru, Pt, Pd, Te, Rh, Ge, Se, Br, Hg, C, B, F, I, H, Mo, O, N, P, S, Sn, W, Sb, Os, Kr, Cl, Ir, Xe, As, Au

Sb :( 25 ): Ru, Pt, Pd, Te, Rh, Se, Br, C, F, I, H, Mo, O, N, P, S, W, Sb, Os, Kr, Cl, Ir, Xe, As, Au

Te :( 22 ): Ru, Pt, Pd, Te, Rh, Se, Br, C, F, I, H, O, N, P, S, Os, Kr, Cl, Ir, Xe, As, Au
I :( 7 ): Cl, I, F, O, N, Kr, Br

Xe :( 7 ): Cl, F, O, N, Kr, Br, Xe

Cs :( 93 ): Ru, Re, Ra, Rb, Rh, Be, Ba, Bi, Bk, Br, H, P, Os, Es, Hg, Ge, Gd, Ga, Pr, Pt, Pu, C, Pb, Pa, Pd, Cd, Po, Pm, Ho, Hf, K, Mg, Mo, Mn, O, S, W, Zn, Eu, Zr, Er, Ni, Na, Nb, Nd, Np, Fe, B, F, Sr, N, Kr, Si, Sn, Sm, V, Sc, Sb, Se, Co, Cm, Cl, Ca, Ce, Xe, Lu, Cs, Cr, Cu, La, Li, Tl, Tm, Th, Ti, Te, Tb, Tc, Ta, Yb, Dy, I, U, Y, Ac, Ag, Ir, Am, Al, As, Au, At, In

Ba :( 89 ): Ru, Re, Rh, Be, Ba, Bi, Bk, Br, H, P, Os, Es, Ge, Gd, Ga, Pr, Pt, Pu, C, Pb, Pa, Pd, Cd, Po, Pm, Ho, Hf, Hg, Mg, Mo, Mn, O, S, W, Zn, Eu, Zr, Er, Ni, Na, Nb, Nd, Np, Fe, B, F, Sr, N, Kr, Si, Sn, Sm, V, Sc, Sb, Se, Co, Cm, Cl, Ca, Ce, Xe, Lu, Cr, Cu, La, Li, Tl, Tm, Th, Ti, Te, Tb, Tc, Ta, Yb, Dy, I, U, Y, Ac, Ag, Ir, Am, Al, As, Au, At, In

La :( 83 ): Pr, Ni, Yb, W, Pt, Ru, Nb, Nd, Mg, Bi, Pb, Y, Tl, Tm, Pd, Ti, Np, Te, Rh, Tc, Pm, Be, Xe, Pa, U, Tb, La, Si, As, Po, Pu, Fe, Br, Dy, Ta, Hf, Hg, Ho, C, B, F, I, H, Mo, Mn, Lu, Bk, O, N, P, S, Re, Kr, Sm, V, Sc, Sb, Es, Os, Se, Th, Zn, Co, Cm, Ag, Cl, At, Ir, Am, Al, Ce, Cd, Ge, Gd, Au, Zr, Ga, In, Eu, Cr, Er, Cu, Sn

Ce :( 61 ): Ni, Pt, Ru, Nb, Mg, Bi, Pb, Y, Tl, Lu, Pd, Ti, Te, Rh, Tc, Ta, Be, Xe, Si, As, Po, Fe, Br, Hf, Hg, At, C, B, F, I, H, Mo, Mn, O, N, P, S, Re, Kr, W, V, Sc, Sb, Os, Se, Zn, Co, Ag, Cl, Ir, Al, Ce, Cd, Ge, Au, Zr, Ga, In, Cr, Cu, Sn

Pr :( 79 ): Pr, Ni, W, Pt, Ru, Nb, Nd, Mg, Bi, Pb, Y, Tl, Tm, Pd, Ti, Np, Te, Rh, Tc, Pm, Be, Xe, Pa, U, Pu, Si, As, Po, Fe, Br, Dy, Ta, Hf, Hg, Ho, C, B, F, I, H, Mo, Mn, Lu, Bk, O, N, P, S, Re, Kr, Sm, V, Sc, Sb, Es, Os, Se, Th, Zn, Co, Cm, Ag, Cl, At, Ir, Am, Al, Cd, Ge, Gd, Au, Zr, Ga, In, Eu, Cr, Er, Cu, Sn

Nd :( 75 ): Ni, W, Pt, Ru, Nb, Nd, Mg, Bi, Pb, Y, Tl, Tm, Pd, Ti, Np, Te, Rh, Tc, Ta, Be, Xe, Pa, U, Pu, Si, As, Po, Fe, Br, Dy, Hf, Hg, Ho, C, B, F, I, H, Mo, Mn, Lu, Bk, O, N, P, S, Re, Kr, Sm, V, Sc, Sb, Os, Se, Es, Zn, Co, Cm, Ag, Cl, At, Ir, Al, Cd, Ge, Gd, Au, Zr, Ga, In, Eu, Cr, Er, Cu, Sn

Pm :( 76 ): Ni, W, Pt, Ru, Nb, Mg, Bi, Pb, Y, Tl, Tm, Pd, Ti, Np, Te, Rh, Tc, Pm, Be, Xe, Pa, U, Pu, Si, As, Po, Fe, Br, Dy, Ta, Hf, Hg, Ho, C, B, F, I, H, Mo, Mn, Lu, Bk, O, N, P, S, Re, Kr, Sm, V, Sc, Sb, Os, Se, Es, Zn, Co, Cm, Ag, Cl, At, Ir, Am, Al, Cd, Ge, Gd, Au, Zr, Ga, In, Eu, Cr, Er, Cu, Sn

Sm :( 74 ): Ni, W, Pt, Ru, Nb, Mg, Bi, Pb, Y, Tl, Tm, Pd, Ti, Np, Te, Rh, Tc, Ta, Be, Xe, Pa, U, Pu, Si, As, Po, Fe, Br, Dy, Hf, Hg, Ho, C, B, F, I, H, Mo, Mn, Lu, Bk, O, N, P, S, Re, Kr, Sm, V, Sc, Sb, Os, Se, Es, Zn, Co, Cm, Ag, Cl, At, Ir, Al, Cd, Ge, Gd, Au, Zr, Ga, In, Eu, Cr, Er, Cu, Sn

Eu :( 68 ): Ni, Pt, Ru, Nb, Mg, Bi, Pb, Y, Tl, Tm, Pd, Ti, Te, Rh, Tc, Ta, Be, Xe, Si, As, Po, Fe, Br, Dy, Hf, Hg, Ho, C, B, F, I, H, Mo, Mn, Bk, O, N, P, S, Re, Kr, W, V, Sc, Sb, Os, Se, Es, Zn, Co, Cm, Ag, Cl, At, Ir, Al, Cd, Ge, Lu, Au, Zr, Ga, In, Eu, Cr, Er, Cu, Sn

Gd :( 70 ): Ni, Pt, Ru, Nb, Mg, Bi, Pb, Y, Tl, Tm, Pd, Ti, Te, Rh, Tc, Ta, Be, Xe, Pa, Si, As, Po, Fe, Br, Dy, Hf, Hg, Ho, C, B, F, I, H, Mo, Mn, Lu, Bk, O, N, P, S, Re, Kr, W, V, Sc, Sb, Os, Se, Es, Zn, Co, Cm, Ag, Cl, At, Ir, Al, Cd, Ge, Gd, Au, Zr, Ga, In, Eu, Cr, Er, Cu, Sn

Tb :( 71 ): Ni, Pt, Ru, Nb, Mg, Bi, Pb, Y, Tl, Tm, Pd, Ti, Te, Rh, Tc, Ta, Be, Xe, Tb, Si, As, Po, Fe, Br, Dy, Hf, Hg, Ho, C, B, F, I, H, Mo, Mn, Bk, O, N, P, S, Re, Kr, W, V, Sc, Sb, Os, Se, Es, Zn, Co, Cm, Ag, Cl, At, Ir, Am, Al, Ce, Cd, Ge, Lu, Au, Zr, Ga, In, Eu, Cr, Er, Cu, Sn

Dy :( 66 ): Ni, Pt, Ru, Nb, Mg, Bi, Pb, Y, Tl, Tm, Pd, Ti, Te, Rh, Tc, Ta, Be, Xe, Si, As, Po, Fe, Br, Dy, Hf, Hg, Ho, C, B, F, I, H, Mo, Mn, O, N, P, S, Re, Kr, W, V, Sc, Sb, Os, Se, Es, Zn, Co, Cm, Ag, Cl, At, Ir, Al, Cd, Ge, Lu, Au, Zr, Ga, In, Cr, Er, Cu, Sn

Ho :( 64 ): Ni, Pt, Ru, Nb, Mg, Bi, Pb, Re, Tl, Tm, Pd, Ti, Te, Rh, Tc, Ta, Be, Xe, Si, As, Po, Fe, Br, Hf, Hg, Ho, C, B, F, I, H, Mo, Mn, O, N, P, S, Kr, W, V, Sc, Sb, Os, Se, Es, Zn, Co, Cm, Ag, Cl, At, Ir, Al, Cd, Ge, Lu, Au, Zr, Ga, In, Cr, Er, Cu, Sn

Er :( 62 ): Ni, Pt, Ru, Nb, Mg, Bi, Pb, Re, Tl, Tm, Pd, Ti, Te, Rh, Tc, Ta, Be, Xe, Si, As, Po, Fe, Br, Hf, Hg, At, C, B, F, I, H, Mo, Mn, O, N, P, S, Kr, W, V, Sc, Sb, Os, Se, Es, Zn, Co, Ag, Cl, Ir, Al, Cd, Ge, Lu, Au, Zr, Ga, In, Cr, Er, Cu, Sn

Tm :( 60 ): Ni, Pt, Ru, Nb, Mg, Bi, Pb, Re, Tl, Tm, Pd, Ti, Te, Rh, Tc, Ta, Be, Xe, Si, As, Po, Fe, Br, Hf, Hg, At, C, B, F, I, H, Mo, Mn, O, N, P, S, Kr, W, V, Sc, Sb, Os, Se, Zn, Co, Ag, Cl, Ir, Al, Cd, Ge, Lu, Au, Zr, Ga, In, Cr, Cu, Sn

Yb :( 75 ): Ni, Pt, Ru, Nb, Mg, Bi, Pb, Y, Tl, Tm, Pd, Ti, Te, Rh, Tc, Ta, Be, Xe, Pa, U, Tb, Yb, Si, As, Po, Fe, Br, Dy, Hf, Hg, Ho, C, B, F, I, H, Mo, Mn, Lu, Bk, O, N, P, S, Re, Kr, W, V, Sc, Sb, Os, Se, Es, Zn, Co, Cm, Ag, Cl, At, Ir, Am, Al, Ce, Cd, Ge, Gd, Au, Zr, Ga, In, Eu, Cr, Er, Cu, Sn

Lu :( 59 ): Ni, Pt, Ru, Nb, Mg, Bi, Pb, Re, Tl, Lu, Pd, Ti, Te, Rh, Tc, Ta, Be, Xe, Si, As, Po, Fe, Br, Hf, Hg, At, C, B, F, I, H, Mo, Mn, O, N, P, S, Kr, W, V, Sc, Sb, Os, Se, Zn, Co, Ag, Cl, Ir, Al, Cd, Ge, Au, Zr, Ga, In, Cr, Cu, Sn

Hf :( 57 ): Ni, Pt, Ru, Nb, Mg, Bi, Pb, Re, Tl, Pd, Ti, Te, Rh, Tc, Ta, Be, Xe, Si, As, Po, Fe, Br, Hf, Hg, C, B, F, I, H, Mo, Mn, O, N, P, S, Kr, W, V, Sc, Sb, Os, Se, Zn, Co, Ag, Cl, Ir, Al, Cd, Ge, Au, At, Ga, In, Cr, Cu, Sn

Ta :( 51 ): Ni, Pt, Ru, Pb, Re, Tl, Pd, Ti, Te, Rh, Tc, Ta, Be, Xe, Si, As, Po, Fe, Br, Hg, C, B, F, I, H, Mo, Mn, O, N, P, S, Kr, W, V, Sb, Os, Se, Zn, Co, Ag, Cl, Ir, Al, Cd, Ge, Au, Ga, In, Cr, Cu, Sn

W :( 13 ): C, W, Cl, I, F, O, N, Kr, S, Au, Br, Xe, Se

Re :( 28 ): Ni, Pt, Ru, Re, Pd, Rh, Tc, As, Br, C, B, P, F, H, O, N, Kr, S, Os, Se, Ag, Cl, Ir, Si, Xe, Ge, Au, Cu

Os :( 17 ): Os, Ru, C, Ir, Pd, Pt, F, H, Cl, O, N, Kr, S, Au, Br, Rh, Se

Ir :( 12 ): C, Ir, Pd, F, H, Cl, O, N, Kr, S, Br, Se

Pt :( 10 ): C, Pt, Cl, F, O, N, Kr, S, Br, Se

Au :( 10 ): C, Cl, F, O, N, Kr, S, Au, Br, Se

Hg :( 24 ): Ru, Pt, Pd, Rh, Ge, Se, Br, Hg, C, B, F, I, H, O, N, P, S, Os, Kr, Cl, Ir, Xe, As, Au

Tl :( 45 ): Ni, Pt, Ru, Pb, Re, Tl, Pd, Te, Rh, Tc, Xe, As, Fe, Br, Hg, C, B, F, I, H, Mo, O, N, P, S, Kr, W, V, Sb, Os, Se, Zn, Co, Ag, Cl, Ir, Si, Cd, Ge, Au, Ga, In, Cr, Cu, Sn

Pb :( 37 ): Ni, Pt, Ru, Pb, Re, Pd, Te, Rh, Tc, As, Br, Hg, C, B, P, F, I, H, Mo, O, N, Kr, S, Sn, W, Sb, Os, Se, Co, Ag, Cl, Ir, Si, Xe, Ge, Au, Cu

Bi :( 28 ): Ru, Pt, Pd, Te, Rh, Bi, Se, Br, C, B, F, I, H, Mo, O, N, P, S, W, Sb, Os, Kr, Cl, Ir, Xe, As, Au, At

Po :( 29 ): Ru, Pt, Pd, Te, Rh, Po, Ge, Se, Br, Hg, C, B, F, I, H, Mo, O, N, P, S, W, Sb, Os, Kr, Cl, Ir, Xe, As, Au

At :( 21 ): Os, Ru, C, Ir, Pd, Pt, F, I, H, Cl, O, N, W, Kr, S, Au, At, Br, Xe, Rh, Se

Ra :( 89 ): Ru, Re, Ra, Rh, Be, Bi, Bk, Br, H, P, Os, Es, Ge, Gd, Ga, Pr, Pt, Pu, C, Pb, Pa, Pd, Cd, Po, Pm, Ho, Hf, Hg, Mg, Mo, Mn, O, S, W, Zn, Eu, Zr, Er, Ni, Na, Nb, Nd, Np, Fe, B, F, Sr, N, Kr, Si, Sn, Sm, V, Sc, Sb, Se, Co, Cm, Cl, Ca, Ce, Xe, Lu, Cr, Cu, La, Li, Tl, Tm, Th, Ti, Te, Tb, Tc, Ta, Yb, Dy, I, U, Y, Ac, Ag, Ir, Am, Al, As, Au, At, In

Ac :( 84 ): Pr, Ni, Yb, W, Pt, Ru, Nb, Nd, Mg, Bi, Pb, Y, Tl, Tm, Pd, Ti, Np, Te, Rh, Tc, Pm, Be, Xe, Pa, U, Tb, La, Si, As, Po, Pu, Fe, Br, Dy, Ta, Hf, Hg, Ho, C, B, F, I, H, Mo, Mn, Lu, Ac, O, N, P, S, Re, Kr, Sm, V, Sc, Sb, Es, Os, Se, Th, Zn, Co, Cm, Ag, Cl, At, Ir, Am, Al, Ce, Cd, Ge, Gd, Au, Zr, Ga, In, Eu, Cr, Er, Bk, Cu, Sn

Th :( 64 ): Ni, Th, Pt, Ru, Nb, Mg, Bi, Pb, Re, Tl, Pa, Pd, Ti, Np, Te, Rh, Tc, Ta, Be, Xe, Si, As, Po, Fe, Br, Hf, Hg, At, C, B, F, I, H, Mo, Mn, Bk, O, N, P, S, U, Kr, W, V, Sc, Sb, Os, Se, Es, Zn, Co, Ag, Cl, Ir, Al, Cd, Ge, Au, Zr, Ga, In, Cr, Cu, Sn

Pa :( 55 ): Ni, Pt, Ru, Nb, Bi, Pb, Re, Tl, Pa, Pd, Ti, Te, Rh, Tc, Ta, Be, Xe, Si, As, Po, Fe, Br, Hg, C, B, F, I, H, Mo, Mn, O, N, P, S, Kr, W, V, Sb, Os, Se, Zn, Co, Ag, Cl, Ir, Al, Cd, Ge, Au, At, Ga, In, Cr, Cu, Sn

U :( 56 ): Ni, Pt, Ru, Nb, Bi, Pb, Re, Tl, Pa, Pd, Ti, Te, Rh, Tc, Ta, Be, Xe, Si, As, Po, Fe, Br, Hg, C, B, F, I, H, Mo, Mn, O, N, P, S, U, Kr, W, V, Sb, Os, Se, Zn, Co, Ag, Cl, Ir, Al, Cd, Ge, Au, At, Ga, In, Cr, Cu, Sn

Np :( 58 ): Ni, Pt, Ru, Nb, Bi, Pb, Re, Tl, Pa, Pd, Ti, Np, Te, Rh, Tc, Ta, Be, Xe, Si, As, Po, Fe, Br, Hg, C, B, F, I, H, Mo, Mn, O, N, P, S, U, Kr, W, V, Sc, Sb, Os, Se, Zn, Co, Ag, Cl, Ir, Al, Cd, Ge, Au, At, Ga, In, Cr, Cu, Sn

Pu :( 65 ): Ni, Pt, Ru, Nb, Mg, Bi, Pb, Re, Tl, Pa, Pd, Ti, Np, Te, Rh, Tc, Ta, Be, Xe, Pu, Si, As, Po, Fe, Br, Hf, Hg, At, C, B, F, I, H, Mo, Mn, Bk, O, N, P, S, U, Kr, W, V, Sc, Sb, Os, Se, Es, Zn, Co, Cm, Ag, Cl, Ir, Al, Cd, Ge, Au, Zr, Ga, In, Cr, Cu, Sn

Am :( 66 ): Ni, Pt, Ru, Nb, Mg, Bi, Pb, Y, Tl, Tm, Pd, Ti, Te, Rh, Tc, Ta, Be, Xe, Si, As, Po, Fe, Br, Hf, Hg, Ho, C, B, F, I, H, Mo, Mn, O, N, P, S, Re, Kr, W, V, Sc, Sb, Os, Se, Es, Zn, Co, Cm, Ag, Cl, At, Ir, Am, Al, Cd, Ge, Lu, Au, Zr, Ga, In, Cr, Er, Cu, Sn

Cm :( 60 ): Ni, Pt, Ru, Nb, Mg, Bi, Pb, Re, Tl, Pd, Ti, Te, Rh, Tc, Ta, Be, Xe, Si, As, Po, Fe, Br, Hf, Hg, At, C, B, F, I, H, Mo, Mn, O, N, P, S, Kr, W, V, Sc, Sb, Os, Se, Es, Zn, Co, Cm, Ag, Cl, Ir, Al, Cd, Ge, Au, Zr, Ga, In, Cr, Cu, Sn

Bk :( 60 ): Ni, Pt, Ru, Nb, Mg, Bi, Pb, Re, Tl, Pd, Ti, Te, Rh, Tc, Ta, Be, Xe, Si, As, Po, Fe, Br, Hf, Hg, At, C, B, F, I, H, Mo, Mn, Bk, O, N, P, S, Kr, W, V, Sc, Sb, Os, Se, Es, Zn, Co, Ag, Cl, Ir, Al, Cd, Ge, Au, Zr, Ga, In, Cr, Cu, Sn

Es :( 59 ): Ni, Pt, Ru, Nb, Mg, Bi, Pb, Re, Tl, Pd, Ti, Te, Rh, Tc, Ta, Be, Xe, Si, As, Po, Fe, Br, Hf, Hg, At, C, B, F, I, H, Mo, Mn, O, N, P, S, Kr, W, V, Sc, Sb, Os, Se, Es, Zn, Co, Ag, Cl, Ir, Al, Cd, Ge, Au, Zr, Ga, In, Cr, Cu, Sn

\textbf{Incomparable bonds.}  The convention is: Bond of element $x$ : (number of incomparable bonds) : Incomparable bonds.

H :( 15 ): C, W, Pt, Cl, I, F, O, N, Kr, S, Au, Br, Xe, Rh, Se

Li :( 47 ): Pr, W, La, Nb, Nd, Pb, Sc, Tl, Tm, Th, Ti, Np, Te, Tb, Po, Pm, Pa, Yb, Bi, Bk, Pu, Ho, Dy, Ta, Hf, Mg, Mo, Lu, U, Sn, Sm, V, Y, Sb, Es, Ac, Cm, Ca, Am, Ce, Cd, Gd, At, In, Eu, Zr, Er

Be :( 43 ): Ni, Pt, Ru, S, Nb, Bi, Pb, Re, Tl, Pd, Te, Rh, Tc, Xe, As, Po, Fe, Br, Hg, I, Mo, Kr, Si, Sn, W, V, Sb, Os, Se, Zn, Co, Ag, P, Ir, Al, Cd, Ge, Au, At, Ga, In, Cr, Cu

B :( 21 ): Ru, Pd, Pt, Cl, I, Os, Mo, Ir, As, Xe, W, P, S, Au, Kr, Br, Sb, Te, Rh, Se, At

C :( 7 ): Cl, I, H, Xe, Kr, S, Br

N :( 2 ): H, Cl

O :( 2 ): H, F

F :( 2 ): H, O

Na :( 28 ): Pr, La, Nd, Pa, Th, Np, Tb, Pm, Yb, Bk, Ho, Dy, Pu, Sr, Lu, U, Sm, Y, Ac, Cm, Ca, Am, Ce, Gd, Eu, Tm, Es, Er

Mg :( 16 ): Nb, Bi, Li, Pb, Np, Tl, Pa, U, Sn, In, Sc, Sb, At, Po, Zr, Ta

Al :( 22 ): Pb, Re, Tl, Cd, Po, Be, Te, Bi, Tc, Hg, I, Mo, Mn, Sn, W, V, Sb, Os, Ag, Xe, At, In

Si :( 24 ): Ru, Pt, Bi, Pd, Te, Rh, Po, Be, Ge, Hg, I, Mo, Kr, Sn, W, Sb, Os, Co, Ag, Ir, Xe, As, Au, At

P :( 17 ): Be, B, Pd, Pt, Ni, I, Os, Ir, Xe, W, Kr, Ru, At, Br, Au, Rh, Se

S :( 8 ): Be, C, B, I, H, Xe, Kr, Br

Cl :( 4 ): H, C, B, N

K :( 1 ): Ra

Ca :( 10 ): Pr, Ac, Th, La, Na, Pu, Nd, Li, Sm, Pm

Sc :( 5 ): Mg, Pa, U, Bi, Li

Ti :( 12 ): Nb, Bi, Li, Pb, Tl, Sn, W, In, Po, Sb, Mo, At

V :( 15 ): Be, Te, Mo, Mn, Al, Cd, Pb, Sn, W, In, Po, Sb, Li, Bi, At

Cr :( 26 ): Ru, Pt, Pb, Re, Cd, Rh, Tc, Be, Te, Bi, Po, Hg, I, Mo, Mn, Sn, W, Sb, Os, Zn, Ag, Xe, Au, At, Ga, In

Mn :( 32 ): Ru, Pt, Nb, Bi, Pb, Re, Tl, Pd, Te, Rh, Tc, Ge, Po, Cd, Hg, I, Mo, Sn, W, V, Sb, Os, Ag, Ir, Al, Xe, As, Au, At, Ga, In, Cr

Fe :( 26 ): Ru, Pt, Bi, Pb, Re, Pd, Te, Rh, Tc, Be, Ge, Po, Hg, I, Mo, Kr, Sn, W, Sb, Os, Ag, Ir, Xe, As, Au, At

Co :( 29 ): Ru, Pt, Bi, Re, Pd, Te, Rh, Tc, Be, As, Po, Br, Hg, I, Mo, Kr, Si, Sn, W, Sb, Os, Se, Ag, Ir, Xe, Ge, Au, At, Cu

Ni :( 26 ): Ru, Pt, Bi, Pd, Te, Rh, Po, Be, Ge, Br, Hg, I, Mo, Kr, Sn, W, Sb, Os, Se, Ag, P, Ir, Xe, As, Au, At

Cu :( 26 ): Ru, Pt, Bi, Pd, Te, Rh, Po, Be, Ge, Br, Hg, I, Mo, Kr, Sn, W, Sb, Os, Se, Co, Ag, Ir, Xe, As, Au, At

Zn :( 29 ): Ru, Pt, Bi, Pb, Re, Pd, Cd, Rh, Tc, Be, Te, Ge, Po, Hg, I, Mo, Sn, W, Sb, Os, Ag, Ir, Xe, As, Au, At, Ga, In, Cr

Ga :( 22 ): Ru, Pb, Re, Te, Rh, Po, Be, Bi, Tc, Hg, I, Mo, Mn, Sn, W, Sb, Os, Zn, Ag, Xe, At, Cr

Ge :( 22 ): Ni, Pt, Ru, Te, Rh, Be, Bi, Fe, I, Mo, Mn, Si, W, Sb, Os, Zn, Co, Ir, Xe, Au, At, Cu

As :( 19 ): Be, Zn, B, Co, Pt, Ni, I, Os, Mn, Xe, Si, Ru, Fe, W, Au, At, Rh, Cu, Ir

Se :( 10 ): Ni, B, Co, Be, I, H, Xe, P, Kr, Cu

Br :( 10 ): Be, C, B, Co, Ni, H, P, S, Kr, Cu

Kr :( 13 ): Be, C, B, Co, Ni, H, S, P, Si, Se, Fe, Br, Cu

Rb :( 0 ): 

Sr :( 2 ): Na, Ac

Y :( 13 ): Pa, Cm, Pu, Na, Li, Tm, U, Ho, Th, Np, Bk, Es, Er

Zr :( 6 ): Mg, Li, Pa, U, Np, Hf

Nb :( 7 ): Be, Mg, Mn, Bi, Li, Ti, Ta

Mo :( 23 ): Ni, Li, Re, Ti, Te, Tc, Be, Fe, Hg, B, Mn, Si, V, Zn, Co, Ag, Al, Cd, Ge, At, Ga, Cr, Cu

Tc :( 20 ): Be, Zn, At, Co, W, I, Bi, Mo, Mn, Al, Xe, Po, Fe, Ga, Sb, Cr, Te, Os, Hg, Sn

Ru :( 17 ): Be, Zn, B, Co, W, Ni, I, Mn, Ge, Xe, P, Si, As, Fe, Ga, Cr, Cu

Rh :( 20 ): Be, Zn, B, Co, Pd, Ni, I, H, Mn, Ge, Xe, W, P, Si, As, Fe, Ga, Cr, Cu, Ir

Pd :( 16 ): Be, Zn, B, Co, Pt, Ni, I, Mn, Xe, P, Si, Au, Fe, W, Rh, Cu

Ag :( 23 ): Ni, Te, Po, Be, Bi, Fe, Hg, I, Mo, Mn, Si, Sn, W, Sb, Os, Zn, Co, Al, Xe, At, Ga, Cr, Cu

Cd :( 16 ): Be, Zn, In, Mo, Mn, Al, Li, Pb, Sn, W, V, Po, Cr, Bi, Sb, At

In :( 14 ): Be, Mg, Mn, Zn, Ti, Al, Li, Pb, At, V, Po, Cr, Cd, Bi

Sn :( 22 ): Ni, Li, Re, Ti, Po, Be, Bi, Tc, Fe, Mg, Mn, Si, V, Zn, Co, Ag, Al, Cd, At, Ga, Cr, Cu

Sb :( 23 ): Ni, Li, Re, Ti, Tc, Be, Fe, Hg, Mg, B, Mn, Si, V, Zn, Co, Ag, Al, Cd, Ge, At, Ga, Cr, Cu

Te :( 22 ): Ni, Li, Re, Tc, Be, Fe, Hg, B, Mo, Mn, Si, W, V, Zn, Co, Ag, Al, Ge, At, Ga, Cr, Cu

I :( 30 ): Ni, Pt, Ru, Re, Pd, Rh, Tc, Be, Si, As, Fe, C, B, H, Mn, P, S, Os, Se, Zn, Co, Ag, Ir, Al, Xe, Ge, Au, Ga, Cr, Cu

Xe :( 29 ): Ni, Pt, Ru, Pd, Rh, Tc, Be, Si, As, Fe, C, B, I, H, Mn, P, S, Os, Se, Zn, Co, Ag, Ir, Al, Ge, Au, Ga, Cr, Cu

Cs :( 0 ): 

Ba :( 1 ): Ra

La :( 3 ): Na, Ca, Li

Ce :( 21 ): Pr, Pa, Th, Cm, Pu, Na, Nd, Am, Li, Np, Gd, U, Ho, Sm, Dy, Eu, Bk, Er, Tm, Es, Pm

Pr :( 6 ): Na, Ca, Yb, Ce, Li, Tb

Nd :( 9 ): Na, Ca, Yb, Am, Ce, Li, Th, Tb, Pm

Pm :( 8 ): Na, Ca, Yb, Nd, Ce, Li, Th, Tb

Sm :( 8 ): Na, Ca, Yb, Am, Ce, Li, Th, Tb

Eu :( 9 ): Pu, Na, Am, Ce, Li, Pa, U, Th, Np
Gd :( 9 ): Pu, Na, Am, Ce, Li, U, Th, Np, Tb
Tb :( 12 ): Pr, Pa, Th, Pu, Na, Nd, Li, Gd, U, Sm, Np, Pm

Dy :( 10 ): Pu, Na, Am, Ce, Li, Pa, U, Th, Np, Bk

Ho :( 10 ): Pu, Na, Ce, Li, Np, Pa, U, Th, Y, Bk

Er :( 11 ): Cm, Pu, Na, Ce, Li, Np, Pa, U, Th, Y, Bk

Tm :( 12 ): Cm, Pu, Na, Ce, Li, Np, Pa, U, Th, Y, Bk, Es

Yb :( 9 ): Pr, Th, Pu, Na, Nd, Li, Sm, Np, Pm

Lu :( 10 ): Cm, Pu, Na, Li, Pa, U, Th, Np, Bk, Es

Hf :( 5 ): Np, Pa, U, Zr, Li

Ta :( 5 ): Bi, Mg, At, Nb, Li

W :( 31 ): Ni, Pt, Ru, Li, Re, Pd, Ti, Te, Rh, Tc, Be, As, Fe, Hg, B, H, Mn, P, Si, V, Os, Zn, Co, Ag, Ir, Al, Cd, Ge, Ga, Cr, Cu

Re :( 18 ): Be, Zn, Co, W, I, At, Mo, Mn, Al, Bi, Po, Fe, Ga, Sb, Cr, Te, Hg, Sn

Os :( 20 ): Be, Zn, B, Co, Ge, Ag, Ni, I, Mn, Al, Xe, W, P, Si, As, Fe, Ga, Cr, Tc, Cu
Ir :( 18 ): Be, Zn, B, Co, Pt, Ni, I, Mn, Ge, Xe, P, Si, As, Au, Fe, W, Rh, Cu

Pt :( 20 ): Be, Zn, B, Co, W, Ni, I, H, Mn, Ge, Xe, P, Si, As, Au, Fe, Pd, Cr, Cu, Ir

Au :( 19 ): Be, Zn, B, Co, Pt, Ni, I, H, Mn, Ge, Xe, P, Si, As, Fe, Pd, Cr, Cu, Ir

Hg :( 20 ): Be, Zn, Co, W, Ag, Ni, Mo, Mn, Al, Bi, Re, Si, Fe, Ga, Sb, Cr, Te, Tc, Cu, At

Tl :( 9 ): Be, Mg, Mn, Al, Li, Po, At, Ti, Bi

Pb :( 16 ): Be, Mg, Mn, In, Zn, Ti, Al, Li, Po, Fe, Ga, V, Cr, Cd, Bi, At

Bi :( 29 ): Ni, Nb, Li, Pb, Re, Tl, Ti, Tc, Ta, Be, Po, Fe, Hg, Mg, Mn, Si, Sn, V, Sc, Zn, Co, Ag, Al, Cd, Ge, Ga, In, Cr, Cu

Po :( 25 ): Ni, Li, Pb, Re, Tl, Ti, Tc, Be, Bi, Fe, Mg, Mn, Si, Sn, V, Zn, Co, Ag, Al, Cd, At, Ga, In, Cr, Cu

At :( 33 ): Ni, Li, Pb, Re, Tl, Ti, Te, Tc, Ta, Be, As, Po, Fe, Hg, Mg, B, Mo, Mn, P, Si, Sn, V, Sb, Zn, Co, Ag, Al, Cd, Ge, Ga, In, Cr, Cu

Ra :( 2 ): K, Ba

Ac :( 4 ): Na, Ca, Sr, Li

Th :( 20 ): Cm, Pu, Na, Ca, Yb, Nd, Am, Ce, Li, Eu, Gd, Ho, Sm, Dy, Y, Tm, Er, Tb, Lu, Pm

Pa :( 19 ): Es, Mg, Hf, Cm, Na, Am, Lu, Eu, Ce, Li, Y, Tm, Ho, Dy, Sc, Bk, Tb, Zr, Er

U :( 20 ): Es, Mg, Hf, Cm, Na, Am, Lu, Eu, Ce, Li, Y, Gd, Ho, Dy, Sc, Tm, Bk, Tb, Zr, Er

Np :( 20 ): Es, Mg, Hf, Cm, Na, Yb, Lu, Am, Ce, Li, Eu, Gd, Ho, Dy, Y, Tm, Bk, Tb, Zr, Er

Pu :( 16 ): Na, Ca, Yb, Lu, Am, Ce, Li, Eu, Gd, Ho, Th, Dy, Y, Tm, Tb, Er

Am :( 14 ): Pa, Th, Pu, Na, Nd, Ce, Li, Np, Gd, U, Sm, Dy, Eu, Bk

Cm :( 12 ): Pa, Na, Lu, Ce, Li, Np, Tm, U, Th, Y, Bk, Er

Bk :( 14 ): Pa, Cm, Na, Lu, Am, Ce, Li, Np, Tm, U, Ho, Dy, Y, Er

Es :( 9 ): Pa, Na, Lu, Ce, Li, Np, Tm, U, Y

\end{document}